\documentclass[format=acmsmall, review=false, screen=true]{acmart}

\usepackage{booktabs} 
\usepackage{makecell}
\usepackage{lipsum}
\usepackage[utf8]{inputenc}
\usepackage{array}
\usepackage{wrapfig}
\usepackage{multirow}
\usepackage{tabularx}
\usepackage{pifont}
\usepackage{lineno}
\usepackage{balance}
\usepackage{xcolor,pifont}

\usepackage{soul}

\usepackage{balance}
\usepackage[linesnumbered, ruled,vlined]{algorithm2e}
\RequirePackage{pdflscape}

\acmJournal{TOSEM}

\usepackage{flushend}
\usepackage{multicol}
\usepackage{placeins}
\usepackage{booktabs}
\usepackage{indentfirst}
\usepackage{fancybox}
\usepackage[most]{tcolorbox}
\usepackage{fontawesome}
\usepackage{mathtools}
\usepackage{MnSymbol,wasysym}
\usepackage{fontawesome}
\usepackage{rotating}
\usepackage{adjustbox}
\usepackage{colortbl}
\usepackage{color}
\usepackage{framed}
\definecolor{Gray}{gray}{0.9}
\definecolor{shadecolor}{gray}{0.95}
\usepackage{tikz}
\usetikzlibrary{trees,positioning,shapes,shadows,arrows}

\tikzset{
  basic/.style  = {draw, text width=2cm, drop shadow, font=\sffamily, rectangle},
  root/.style   = {basic, rounded corners=2pt, thin, align=center, fill=white},
  level-2/.style = {basic, rounded corners=6pt, thin,align=center, fill=white, text width=3cm},
  level-3/.style = {basic, thin, align=center, fill=white, text width=1.8cm}
}
\usepackage{tcolorbox}

\newcommand{\todo}[1]{}
\renewcommand{\todo}[1]{{\color{red} TODO: {#1}}}

\begin{document}
%
\title{Accessibility in Software Practice:  A Practitioner's Perspective}

	\author{Tingting Bi}

	\affiliation{%
		\institution{the Faculty of Information Technology, Monash University, Melbourne, Australia.}
	}
	\email{tingting.bi@monash.edu}
    
 	\author{Xin Xia}
	\affiliation{%
		\institution{the Faculty of Information Technology, Monash University, Melbourne, Australia.}
	}
	\email{xin.xia@monash.edu}   
    
    \author{David Lo}
	\affiliation{%
		\institution{the School of Information Systems, Singapore Management University, Singapore.}
	}
	\email{davidlo@smu.edu.sg}
    
    \author{John Grundy}

	\affiliation{%
		\institution{the Faculty of Information Technology, Monash University, Melbourne, Australia.}
	}
	\email{john.grundy@monash.edu}
    
      \author{Thomas Zimmermann}

	\affiliation{%
		\institution{Microsoft Research, Redmond, WA, USA.}
	}
	\email{tzimmer@microsoft.com}

     \author{Denae Ford}

	\affiliation{%
		\institution{Microsoft Research, Redmond, WA, USA.}
	}
	\email{denae@microsoft.com}



%



\begin{abstract}
Being able to access software in daily life is vital for everyone, and thus accessibility is a fundamental challenge for software development. However, given the number of accessibility issues reported by many users, e.g., in app reviews, it is not clear if accessibility is widely integrated into current software projects and how software projects address accessibility issues. In this paper, we report a study of the critical challenges and benefits of incorporating accessibility into software development and design. We applied a mixed qualitative and quantitative approach for gathering data from 15 interviews and 365 survey respondents from 26 countries across five continents to understand how practitioners perceive accessibility development and design in practice. We got 44 statements grouped into eight topics on accessibility from practitioners' viewpoints and different software development stages. Our statistical analysis reveals substantial gaps between groups, e.g., practitioners have Direct v.s. Indirect accessibility relevant work experience when they reviewed the summarized statements. These gaps might hinder the quality of accessibility development and design, and we use our findings to establish a set of guidelines to help practitioners be aware of accessibility challenges and benefit factors. We also propose some remedies to resolve the gaps and to highlight key future research directions.  
\end{abstract}
\maketitle

\keywords{
Accessibility Development and Design, Challenges, Empirical Study, Practitioner
}



\section{Introduction}

Software systems are becoming increasingly complex, and the increasing need to use them in both work and daily life makes users ever more dependent on them. \textit{Accessibility} is one of the prominent software qualities that determines usability and acceptance of a software product in today's competitive market \cite{henss2012semi}\cite{e2019accessible}. An accessible information technology solution that is usable by people with a wide range of differences exhibits "universal design" \cite{burgstahler2011universal}. Universal design is a collection of concepts and approaches that can be applied and incorporated into a product so that end-users can easily access software applications \cite{king2009universal}. Most software systems assume that users can efficiently perform all the following tasks: read the text and images displayed on the screen and respond to them; use the keyboard for typing; use the mouse for selecting text or other information; hear the sounds and react to them, and manipulate items using the mouse, finger or gesture \cite{brunet2005accessibility}. However, many groups of people have difficulties performing one or more of the above tasks due to physical and/or mental challenges that prevent them from using many popular software applications. \cite{lisney2013museums}. 

The importance of accessibility is increasingly being recognized. For example, many countries have mandated accessibility laws and policies for organizations\footnote{https://www.w3.org/WAI/policies/}. In addition, for an application to be successful, it is not enough to satisfy all functional requirements that are expected of it. Accessibility needs to be considered at all stages of the application development life cycle, for example, requirements specification, design, implementation, and testing \cite{yusop2016reporting}. Nevertheless, it is not clear how accessibility features are currently incorporated into the software engineering process and current software products \cite{vendome2019can}. The initial omission of accessibility features often results in drawbacks, such as poor usability, poor user feedback, and considerable re-engineering. Furthermore, accessibility requirements have evolved -- accessibility is now more likely to ensure a wider range of end-users can use their software efficiently and effectively \cite{tidwell2010designing}. 

A variety of research focuses on understanding accessibility development and design for specific disabilities \cite{archambault2008towards,zhou2020making}. For example, previous works have investigated accessibility issues in Android apps \cite{alshayban2020accessibility,vendome2019can}. Whilst these works are valuable for understanding accessibility issues, there is a paucity of empirical evidence on how practitioners perceive accessibility development and design in practice, and how projects go about providing accessibility features for different end-users. A better understanding on the characteristics of accessibility and software practitioners' opinions would provide more tailored support for accessibility design and implementation. The following gaps illustrate that accessibility needs to be considered in software development and design:

\begin{itemize}
\item Accessibility issues are often addressed in the later development stages \cite{curtis2008planning}. Both the Human Computer Interaction (HCI) and Software Engineering (SE) communities report that accessibility plays a crucial role in development \cite{sears2011representing,  condori2018characterizing}. Accessibility is however sometimes not a requirement in the first place when developing the projects.

\item Difficulties in incorporating accessibility into projects. Special knowledge on disabilities and standards about accessibility features is required \cite{britto2016towards}. Trade-offs of incorporating and managing accessibility during the software life cycle are hard to address \cite{roh2017ontological}.

\item Accessibility is essential for both general and specific end-users. Accessibility should be considered throughout the whole software development life cycle. Most importantly, accessibility that is not well managed (e.g., remaining implicit or becoming late requirements) can lead to severe software development issues \cite{sanchez2017method}.  

\end{itemize}
Software engineering plays a fundamental role in developing accessible applications since it promotes the integration between methodologies and specific accessibility techniques and activities in the software development process \cite{paiva2020accessibility}. Given the importance of accessibility, we aim to determine accessibility development and design in software practice. As much of this field has not been grounded in the practitioners' perspective, we conducted a mixed quantitative and qualitative study to gather data from their viewpoint. In this paper, we advocate that accessibility needs to be treated as a first-class consideration throughout software development, i.e., accessibility considerations are critical for improving end-users use of software products. This is also corroborated by other studies \cite{alshayban2020accessibility, shinohara2018teaches}. The main contributions of this work include:

\begin{itemize}
\item We investigate how practitioners perceive accessibility design and development in practice. To the best of our knowledge, this is the first attempt to investigate how prevalent accessibility needs are incorporated into software projects from the participants' perspectives. We present results of 15 detailed interviews and an online survey with 365 responses from 26 countries. 
\item Our analysis of our interviews and online survey show that accessibility is a prevalent consideration in software practice. However, it is often addressed in a simple way and with short term goals. In addition, organizational factors strongly impact the success of accessibility for many projects.
\item We highlight empirical evidence regarding how accessibility fits into software project considerations, and identify a set of challenges that the projects face when addressing these accessibility issues. We provide a set of guidelines to help practitioners be aware of these challenges. We also list suggestions for incorporating accessibility considerations into projects.
\item We identify a set of gaps between work experiences of different participants, e.g., between Web and Mobile App development practitioners on how they perceive accessibility. We attempt to form a better understanding of accessibility challenges in different domains and contexts.

\end{itemize}
The remainder of this paper is structured as follows: Section \ref{Section 3 Research Methodology} describes our study. Section \ref{Section 4 Results} and Section  \ref{Section 5 Discussion} answers and discusses the results of our study, respectively. Section \ref{Section 6 Threats to validation} presents the threats to validity. Section \ref{Section 2 Background and Related work} introduces related works. Finally, Section \ref{Section 7 Conclusion} concludes this work with future directions.

\section{Study Design}
\label{Section 3 Research Methodology}

\subsection{Objective and research questions}
In this work, we aim to help practitioners better understand and identify accessibility challenges and issues in practice. Following this aim, we guided our study with two research questions:

\textbf{RQ1: What are practitioners' perceptions of accessibility development and design in practice?}

\textbf{Motivation}: To answer this RQ, we planned to explore how practitioners perceive accessibility development and design in their software development and design practices. We explore this question from the following perspectives: 
\begin{itemize}
\item \textbf{Understanding accessibility}: What level do practitioners consider the importance of accessibility and what motivates them to incorporate and address accessibility issues in practice? 
\item \textbf{Work characteristics}: How do practitioners' work characteristics impact accessibility development and design in practice? For example, (1) \textit{Skill variety}: Does accessibility design and development intensively require specific knowledge? (2) \textit{Task complexity and problem-solving}: Does software design and development focus on specific software aspects to enhance accessibility? (3) \textit{Task identification}: Is it harder to make an accurate plan for accessibility-related design tasks? (4) \textit{Interaction}: To incorporate accessibility considerations, do development teams have to communicate more frequently with their clients?
\end{itemize}
\begin{itemize}
\item \textbf{Organizational factors}: Whilst software functional requirements and quality requirements are well defined in shaping software design, other factors, such as company culture, development team size, and development platform, may impact the ability of developers to incorporate accessibility into a system \cite{bi2018architecture}. We plan to investigate what organizational factors, in participations' opinions, would impact accessibility development and design.
\end{itemize}

\textbf{RQ2}\textbf{: How does addressing accessibility needs fit into the software development life cycle?}

\textbf{Motivation}: To answer this RQ, we wanted to systemically investigate how accessibility fits into the software development life cycle from a practitioner's perspective, including requirements elicitation, software design, implementation, testing, and evaluation. Answers to this RQ can help us to characterize current accessibility design approaches, limitations, and highlight future work directions:
\begin{itemize}
\item \textbf{Requirements for accessibility}: Collecting requirements to incorporate accessibility involves more preliminary efforts. We plan to understand what challenges practitioners have met when they elicit accessibility-related requirements.
\item \textbf{Design and implementation for accessibility}: Detailed design incorporating accessibility may be more time-consuming and tends to be conducted in an intensively iterative way. What issues developers have met when they are dealing with accessibility design and potential solutions that have been proposed?
\item \textbf{Testing and quality assurance for accessibility}: Quality assurance is important in any projects. However, good performance during testing can not guarantee accessibility of a system to a diverse range of users. Furthermore, it is hard to involve a sufficiently broad end-user base for conducting comprehensive accessibility testing.
\item \textbf{Evaluation for accessibility}: Accessibility requirements drive relevant functionality implementation. It is also necessary to consider when making trade-offs to implement accessibility requirements. How do practitioners evaluate the process of software development to incorporate accessibility into the system?
\end{itemize}

\subsection{Methodology}

\begin{figure*}
\centering
\includegraphics[width=\textwidth]{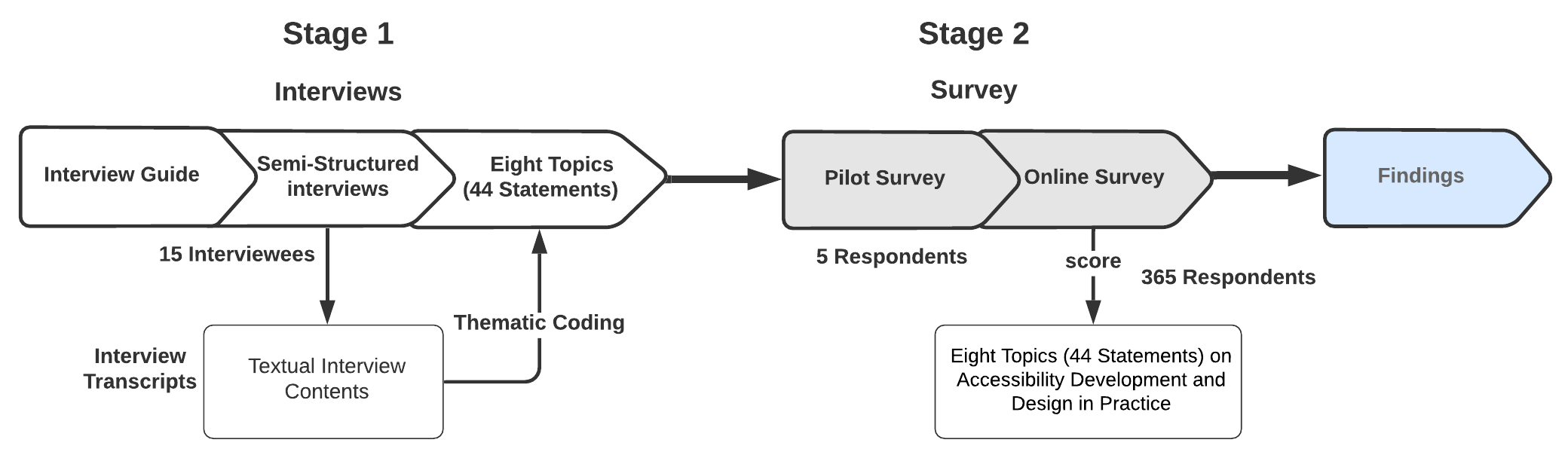}
\caption{Sequential mixed-methods approach includes semi-structured interviews and an online survey.}
\label{Fig_Research methodology}
\end{figure*}

Our research methodology, followed a mixed qualitative and quantitative approach \cite{easterbrook2008selecting}, as depicted in Fig. \ref{Fig_Research methodology}. In the first stage, we conducted open-ended and detailed interviews with 15 software practitioners to get their opinions on accessibility development and design in practice. In the second stage, we conducted an online survey, which received 365 validated responses, to confirm or refute the results of the interviews\footnote{The number of our Human Research Ethics Committee approval from Monash University is 24733.}. We present detailed procedures of each stage in Section \ref{interviews} and Section \ref{survey}, respectively.

\subsubsection{Interviews}
\label{interviews}

\textbf{Participant selection}. We recruited full-time software practitioners from three IT companies in China and the U.S., namely, Alibaba, Hengtian, and Microsoft. Interviewees were recruited by emailing our personal contacts in each company or project, who disseminated the news of our study to their colleagues. Volunteers would inform us if they were willing to participate in the study with no compensation. With this approach, we included 15 interviewees in this study. The 15 interviewees specialize in accessibility design, and they have varied job roles with 9 years of professional experience on average (min 4 years and max 15 years). In the remainder of this paper, we refer to these 15 interviewees as I1 to I15. To preserve the anonymity of participants, we anonymized all items that constitute Personally Identifiable Information (PII) before analyzing the data \cite{mccallister2010guide}. We summarized information about our 15 interviewees in Table \ref{Table_interviewees}.

\begin{table}
\small
\caption{Interviewees and their particular roles.}
\label{Table_interviewees}
\centering
\begin{tabular}{c|c}
\hline
\textbf{Main Role} & \textbf{Interviewee Codes}\\
\hline
\hline
Programming & I1, I2, I6, I9, I12, I13, and I14 \\
\hline
Design & I3, I4, I5, and I10 \\
\hline
Management & I7, I10, and I11 \\
\hline
Testing & I2, I6, I8, and I9 \\
\hline
\end{tabular}
\end{table}

\textbf{Interviewing Process}. The first author conducted a series of interviews with the 15 interviewees, and each interview was completed within 30 minutes. The interviews were semi-structured and divided into three parts.

\textit{Part 1}: We asked some demographic questions, such as the interviewees' experience in software development /testing /project management. We asked interviewees to describe the projects they have done as well as their development team information.

\textit{Part 2}: We asked open-ended questions to understand their opinions on software design and development for accessibility. The questions include: (1) their general opinions on accessibility (e.g., concepts of accessibility in practice); (2) accessibility use cases; (3) requirements elicitation used to inform accessibility design; (4) software design and implementation for accessibility; (5) testing and evaluation for accessibility; (6) tools and standards they have applied or followed for accessibility.

\textit{Part 3}: We prepared candidate topics by carefully reading the contents of representative textbooks. We picked a list of topics that have not been explicitly mentioned in the open discussion and asked the interviewees to discuss those topics further. At the end of each interview, we thanked the interviewees and briefly informed them what we plan to do with his/her response.

\textbf{Transcribing and coding}. We used a commercial transcription service provided by a third-party company to transcribe recordings to transcripts. We then read the transcripts and conducted a thematic coding analysis of the transcripts \cite{clarke2015thematic}. We dropped sentences during the coding process that are not related to "software design and development for accessibility in practice". The first two authors read and coded the contents of transcripts. We used the MAXQDA\footnote{https://www.maxqda.com/} tool for analyzing and coding the qualitative data. To ensure the quality of codes, we invited a Ph.D. candidate to verify the first two authors' initial codes and provided suggestions for improvement. After incorporating these suggestions, we generated a total of 288 cards that contain the codes. After merging the codes with the same words or meaning, we had a total of 122 unique codes. We noticed that when our interview transcripts reached saturation and new codes did not appear anymore, our list of codes was then considered stable.

\textbf{Data analysis and triangulation}. Textual analysis of the qualitative data collected from the interviews was performed primarily using Thematic Analysis \cite{clarke2015thematic}, combined with open coding with the identified themes as underlying codes emerged. The first two authors separately analyzed the codes and sorted the generated cards into potential themes. Themes were not chose before the analysis. We then used Cohen's Kappa \cite{cohen1960coefficient} measure to examine the agreement between the annotators, and the overall Kappa value is 0.79, which indicates substantial agreement between the two annotators. To reduce bias from two annotators sorting the cards to form initial themes, they reviewed and agreed on the final set of themes. Finally, we summarized 44 statements grouped into eight topics about accessibility from a variety of perspectives and across different development phases (the detailed results are shown in Table \ref{Table2_Statements} of Section \ref{Results for RQ1}).

\subsubsection{Survey}
\label{survey}
We then conducted an online survey aiming at confirming, refuting, and extending our results from the 15 interviews. We followed Kitchenham and Pfleeger's guideline for personal opinion surveys \cite{kitchenham2008personal} and used an anonymous survey to increase the response rate \cite{tyagi1989effects}\cite{morrel2002getting}. 

\textbf{Survey design}. Our survey includes different types of questions, i.e., multiple choice and free-text answer questions. We first piloted the preliminary survey with a small set of practitioners (i.e., 5 participants) who came from Alibaba. We obtained the feedback on (1) whether the length of the survey was appropriate, and (2) the clarity and understandability of the terms. We made minor modifications to the draft of the survey based on the received feedback and produced a final version. Note that the collected responses from the pilot survey are excluded from the presented results in this paper. The first author translated the survey into Chinese and this was checked for accuracy by the second author. Our questionnaire can be found in the following links\footnote{English version: https://tinyurl.com/yymexzo9 
    Chinese version: https://www.wjx.cn/jq/93215056.aspx}.

The survey consists of three sections: (1) Demographic information; (2) Statement scoring; (3) and Rationale and Suggestions on "software design for accessibility". 
\begin{itemize}
\item Demographics. We collected geographical location, the respondent's role, size of the team the respondent is in or manages, number of years the respondent has been in their current role, and accessibility work experience, and type of applications that they have developed. We designed two sets of questions about their accessibility working experience. Apart from the basic demographic questions, we designed two sets of questions in this section. \textbf{Question set 1}: "\textit{Have you ever done any software activities (e.g., design, coding, and testing) supporting software accessibility}?"; "\textit{Has your development team carried out any accessibility tasks?}", and "\textit{Please briefly describe how familiar are you with software accessibility}\textit{?}". \textbf{Question set 2}: "\textit{In which domain, have you done development and design for accessibility? Can you please briefly describe the project?}". 
\item Statement scoring. In this section, based on the results of interviews, we developed 44 statements grouped into eight topics, and we provided online survey participants with those statements. Participants were asked to score the statements according to their work experience and opinions. They assessed the importance of each statement on a 5 point Likert scale (i.e., level of the agreement) and one more extra option (Strongly Disagree, Disagree, Neutral, Agree, Strongly Agree, and I Don't Know). The "I Don't Know" option was provided if some statements are not applicable to respondents' experience or for respondents who had a poor understanding of the statement.
\item Rationale and suggestions. Apart from scoring the statements, they also can provide the rationale and suggestions for each statement and topic.
\end{itemize}

\textbf{Recruitment of Respondents}. To recruit respondents from the population of practitioners, we distributed the survey to a broad range of companies from various locations worldwide. No identifying information was required or gathered from our respondents. To get a sufficient number of respondents from diverse background, we followed a multi-pronged strategy to recruit respondents:
\begin{itemize}
\item We contacted professionals from a set of companies worldwide and asked them to disseminate our survey. Specifically, we sent emails to our contacts at Alibaba, Baidu, Hengtian, IBM, Microsoft, and other companies. We adopted a snowballing method to encourage them to disseminate our survey to some of their colleagues willing to participate in our survey. 
\item We sent an email with a link to the survey to 3,987 developers from GitHub projects. We aimed to recruit open-source developers working in the software industry.
\end{itemize}

\textbf{Data analysis of the survey}. In total, we got 431 responses, and we excluded responses who had no direct or indirect (definitions are listed below) accessibility-related work experience. Finally, we had 365 valid responses for analysis. The 365 respondents reside in 26 countries across five continents, and the top countries in which the respondents reside are China (177 respondents) and the United State (105 respondents). The majority of our responses came from participants with two groups: Web App development (181 respondents) and Mobile App development (172 respondents). We analyzed the survey results based on the question types. For multiple-choice questions, we report the percentage of each option  selected. To identify the agreement level of each statement, we analyzed the Likert-scale ratings. Furthermore, we extracted comments that our survey respondents explain why they give such a score of a particular statement.

\textbf{Comparison}. To better understand participants' opinions on accessibility, we divided all survey respondents into different demographic groups and compared their scoring results towards various challenges and desired improvements mentioned by interviewees. We considered the following demographic groups, note that for the size of the teams, we followed the previous work \cite{kochhar2016practitioners}, which defines a group with less than 20 developers as a relatively small team: 

\begin{itemize}
\item Respondents with direct accessibility work experience (108 respondents).
\item Respondents with indirect accessibility work experience (i.e., their groups have done accessibility relevant work, but the respondents have not been directly involved in relevant tasks) (257 respondents).
\item Respondents who are working in a relatively big team ($\geq$ 20) (103 responses).
\item Respondents who are working in a relatively small team (\textless  20) (262 responses).
\item Respondents who mainly develop Web applications (181 responses). 
\item Respondents who mainly develop Mobile  applications (172 responses). 
\end{itemize}

We highlighted the statements with statistically significant difference using different symbols in results for RQ1 and RQ2 (Section \ref{Results for RQ1} and \ref{results of RQ2}). We also summarized the gaps between the demographic groups in Section \ref{Gap differ across groups}.

\begin{figure}
\centering
\label{Fig_Domains}
\includegraphics[width= 7.5 cm]{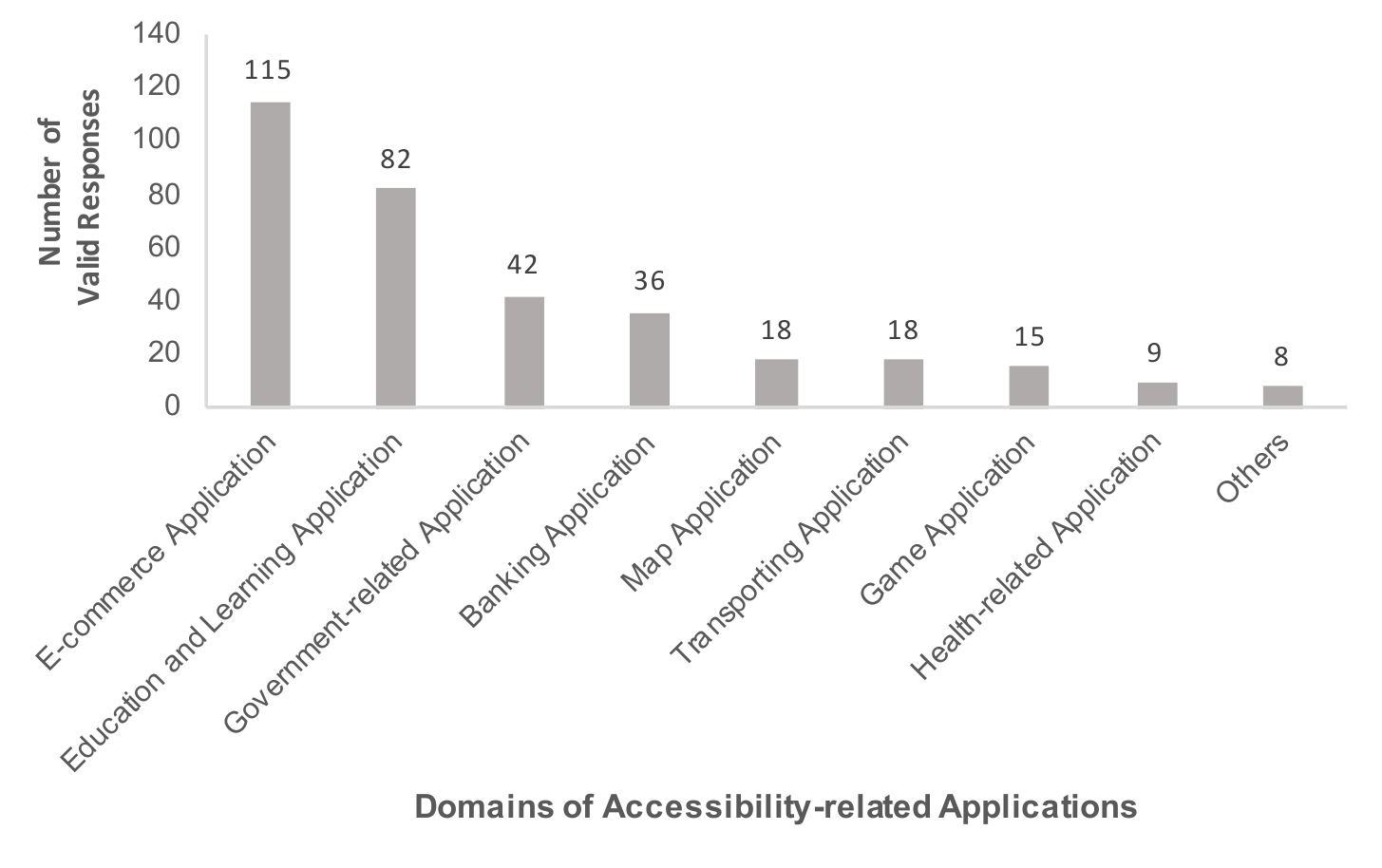}
\caption{Project domains of incorporating accessibility that identified by practitioners of the online survey. }
\label{Fig_Software demographics}
\end{figure}

The statements and the results of the online survey are summarized in Table 2. We analyzed the Likert scale "Agreement" to each statement, and P-value is applied to test whether the discrepancies \textit{in the level of} "Agreement" for each statement are statistically significant differences between the two groups at a 95$\%$ confidence level. We conducted Bonferonni correction for the multiple comparisons \cite{armstrong2014use}. The results show that 15 statements with statistically significant discrepancies between the groups are highlighted with different color cells (the detailed colors are explained below).

The Effect Size \cite{murphy2014cowboys, 9263357} quantifies the difference between groups, and the values of Effect Size are shown in Table 2. For example, the mean score of \textit{all} the respondents for S1 is 4.25. The mean score of for practitioners from \textbf{Indirect} accessibility relevant working experience group is 4.18, whereas the mean source for practitioners from \textbf{Direct} accessibility related working experience is 4.36, as a consequence, the effect size is 4.18 - 4.36 = - 0.18, which means participant group with \textbf{Indirect} accessibility relevant work experience agrees this statement more. We use a Dark Grey color to indicate the former group is more likely to agree with the statement, and Light Grey color indicates the latter group is more likely to agree with the statement.

We numbered statements in the order, depending on the topics that we summarized, in which they appeared in the survey. We annotated each with whether they are statistically significant or not as follows:

\faWrench \ indicates significant difference between participant groups with \textbf{Direct} accessibility work experience v.s. respondents with \textbf{Indirect} accessibility work experience. 

\phone \ indicates significant difference between participant groups who develop on \textbf{Web} v.s. the \textbf{Mobile} applications.

\faGroup \ indicates significant difference between participant groups who work at \textbf{Big} v.s. \textbf{Small} teams.

For example, [\textbf{Sx}] means Statement x has no significant differences between groups;  [\faWrench \ \phone \ \textbf{Sx}] means there is a significant difference between participant groups with Direction and Indirection accessibility relevant work experience as well as Web App and Mobile App development groups regarding perceiving statement x.

\section{Study Results}
\label{Section 4 Results}

We report our findings for RQ1 (Section \ref{Results for RQ1}) and RQ2 (Section \ref{results of RQ2}). We summarize several topics identified by using open card sorting on the interview contents, and each topic includes a set of statements. We provide the percentage of agreements and disagreements for each statement from the online survey. We also selected some comments and highlighted relevant statistics that we derived from our survey responses. We report the differences across different groups regarding statements in Section \ref{Gap differ across groups}.

We applied \faThumbsOUp  \ to indicate \textit{positive} (i.e., benefits) comments and \faThumbsODown  \ to indicate \textit{negative} (i.e., challenges) comments when considering/incorporating accessibility into a project.

\subsection{RQ1: What are practitioners' perceptions of accessibility development and design in practice?}

We answer RQ1 from three aspects, i.e., how practitioners understand accessibility (see Section \ref{SubsecRQ1_ Understanding}), how work characteristics impact accessibility (see Section \ref{SubsecRQ1_ Work}), and how organizational factors impact accessibility design and development in practice (see Section \ref{SubsecRQ1_ Organizational}).

\label{Results for RQ1}








\definecolor{cellorange}{rgb}{ 1,  .949,  .8}
\definecolor{cellgreen}{rgb}{ .776,  .878,  .706}
\definecolor{cellblue}{rgb}{ .608,  .761,  .902}
\definecolor{celllightgrey}{rgb}{ .921,  .921,  .921}
\definecolor{celldarkgrey}{rgb}{ .664,  .664,  .664}

\definecolor{shadecolor}{rgb}{.92,  .92, .92}


\begin{landscape}

\begin{table}

  \centering

 \caption{ Interview and Survey results on accessibility statements.  \faWrench \ indicates statistically significant differences between \textbf{Indirect} and \textbf{Direct} work experience groups. \faGroup \ represents statistically significant differences between \textbf{Big} and \textbf{Small} size groups. \phone \ represents statistically significant differences between \textbf{Web} and \textbf{App} development groups.  \colorbox{celldarkgrey}{Dark grey} cells indicate where former agrees more.  \colorbox{celllightgrey}{Light grey} cells indicate where the latter group agree more. The number in the Likert Distribution column indicates the size of each group. The bars in the Likert distributions from left to right are: Strongly Disagree (1 score), Disagree (2 scores), Neutral (3 scores), Agree (4 scores), Strongly Agree (5 scores), and I Don't Know option.}

 \label{Table2_Statements}
\tiny

\begin{adjustbox}{totalheight={12.5 cm}}
    \begin{tabular}{r|c|cc|rr|rr|rr}

    \hline

             &  & \multicolumn{2}{c|}{\textbf{Likert Distributions}} & \multicolumn{2}{c|} {\textbf{Indirect v.s. Direct}} & \multicolumn{2}{c|}{\textbf{Big v.s. Small}} & \multicolumn{2}{c}{\textbf{Web v.s. App}}  \\    
          
 \textbf{Statement} &  & In total & Score & P-value & Effect Size & P-value & Effect Size & P-value & Effect Size\\

\specialrule{0.1em}{0pt}{0.5pt}
\specialrule{0.1em}{0pt}{0.5pt}

\multicolumn{10}{l}{\textbf{T1. General Considerations of Accessibility}}\\

\hline

Accessibility needs to be incorporated into all software projects. & S1 & \includegraphics[width = 0.4cm, height = 0.12  cm]{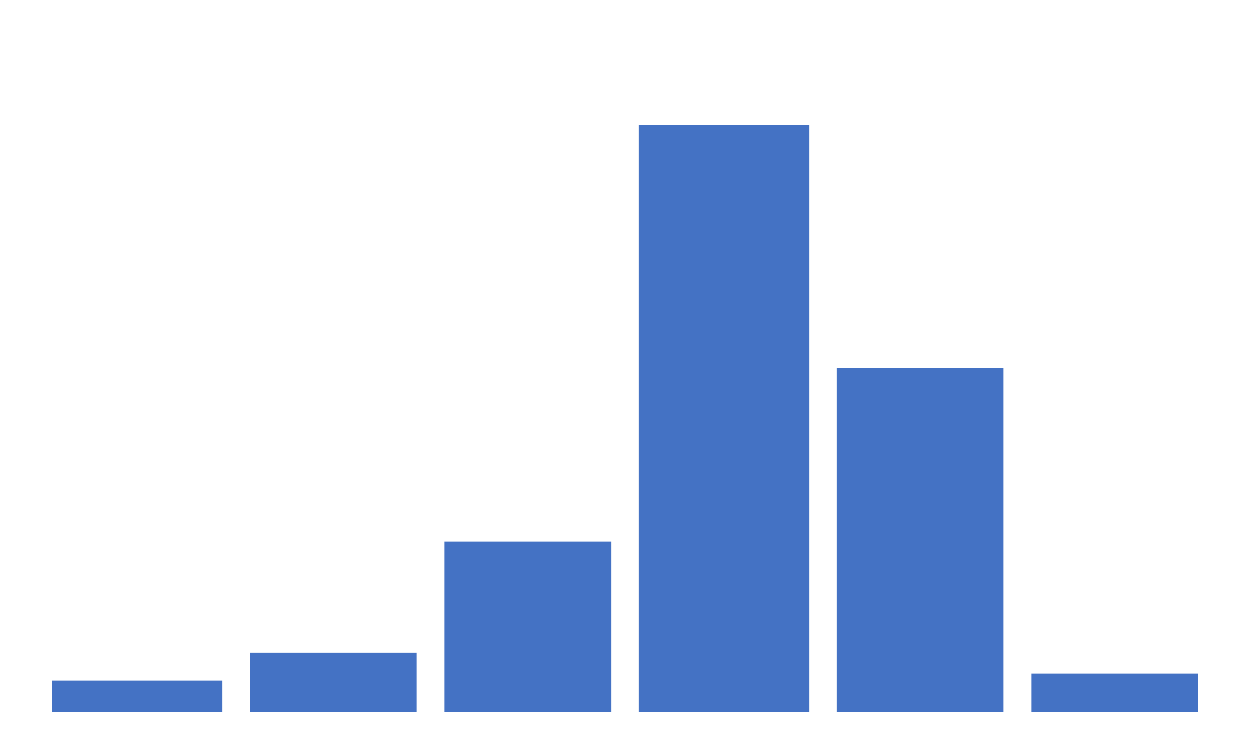} & 4.25 & \faWrench \ \textbf{.000} & \cellcolor[rgb]{ .921,  .921,  .921}-0.18 & .947 & 0.02 & .306 & 0.15 \\

 Accessibility is not only for people who are unable to use standard software.  & S2 & \includegraphics[width = 0.4cm, height = 0.12  cm]{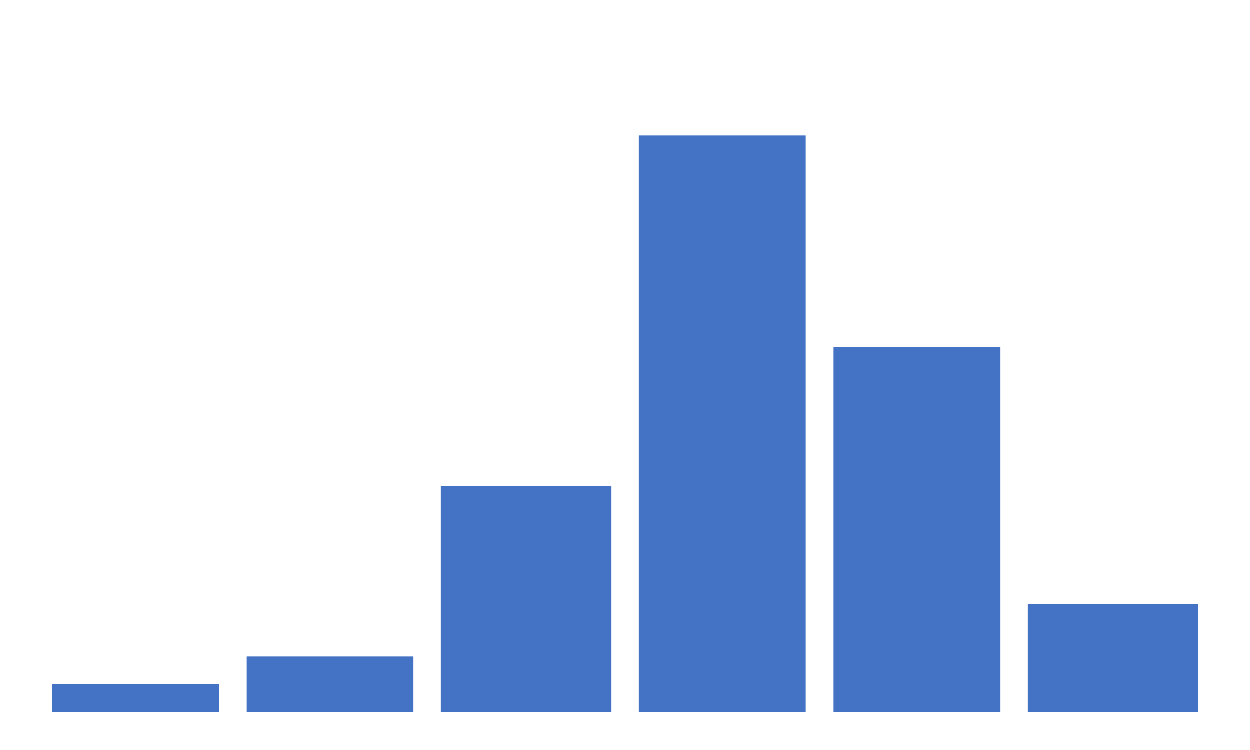} & 4.19 & .345 & 
-0.28 & .244 & -0.33 & .890 & 0.23  \\

 Accessibility should be integrated with all software activities.  & S3 & \includegraphics[width = 0.4cm, height = 0.12  cm]{Bigtable_figures/Concept_2.pdf} & 4.19 & .345 & 
-0.28 & .244 & -0.33 & .890 & 0.23  \\

 Accessibility needs evolve during software development and design. & S4 & \includegraphics[width = 0.4cm, height = 0.12  cm]{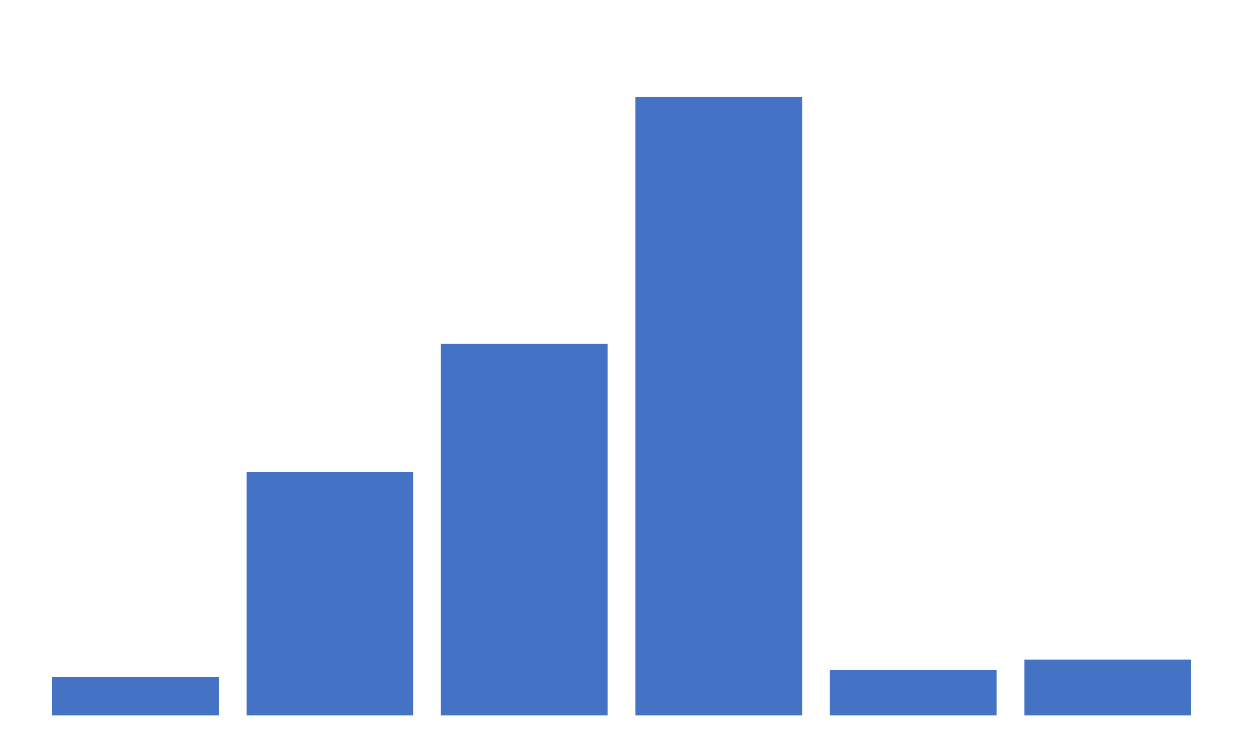}  & 4.01 & 1.00 & 
0.09 & .806 & 0.22 & \phone \ \textbf{.000} & \cellcolor[rgb]{  .921,  .921,  .921}-0.38  \\

 Goals for accessibility: easy to read, easy to operate, and simple to use. & S5 & \includegraphics[width = 0.4cm, height = 0.12  cm]{Bigtable_figures/Concept_3.pdf}  & 4.01 & 1.00 & 
0.09 & .806 & 0.22 & .079 & 0.38  \\

 Accessibility design drives innovation  and often solves unanticipated problems.  & S6 & \includegraphics[width = 0.4cm, height = 0.12  cm]{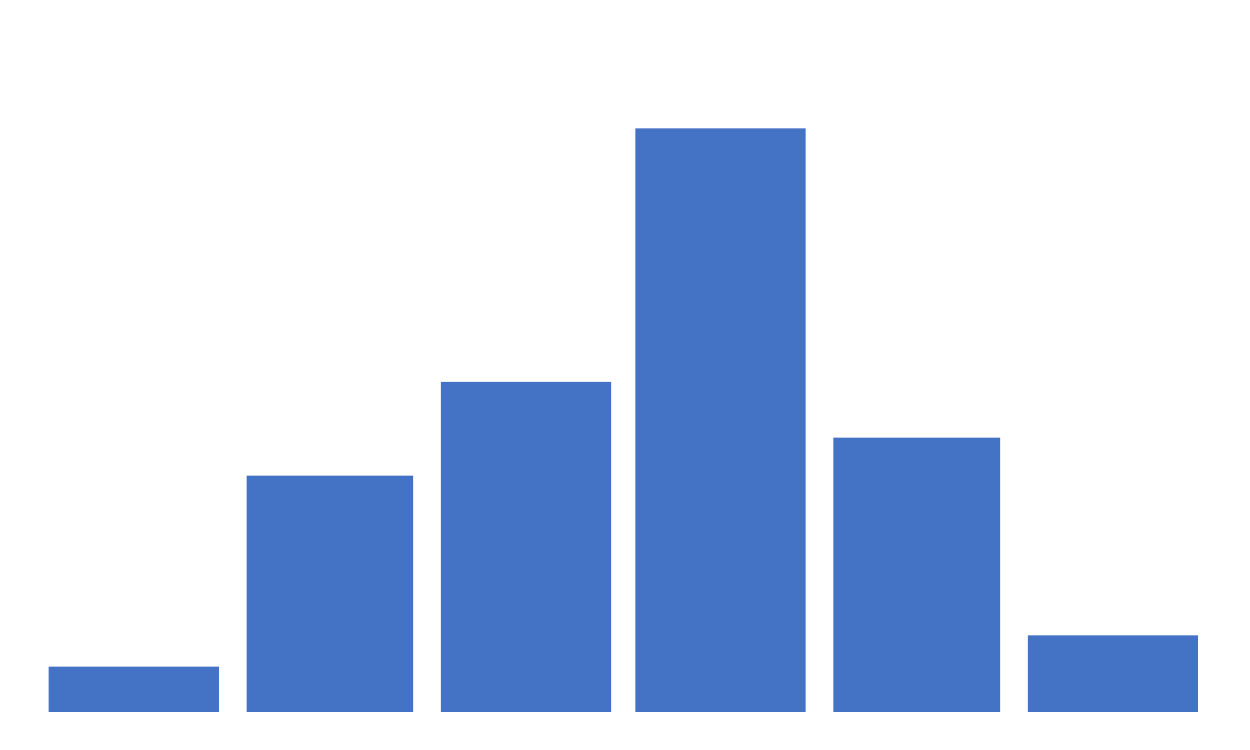}  &  3.92 & 1.00 & 
0.17 & .265 & -0.16 & \phone \ \textbf{.001} &\cellcolor[rgb]{ .664,  .664,  .664}  0.38  \\

 Accessibility is a widely considered concept in software development. & S7 & \includegraphics[width = 0.4cm, height = 0.12  cm]{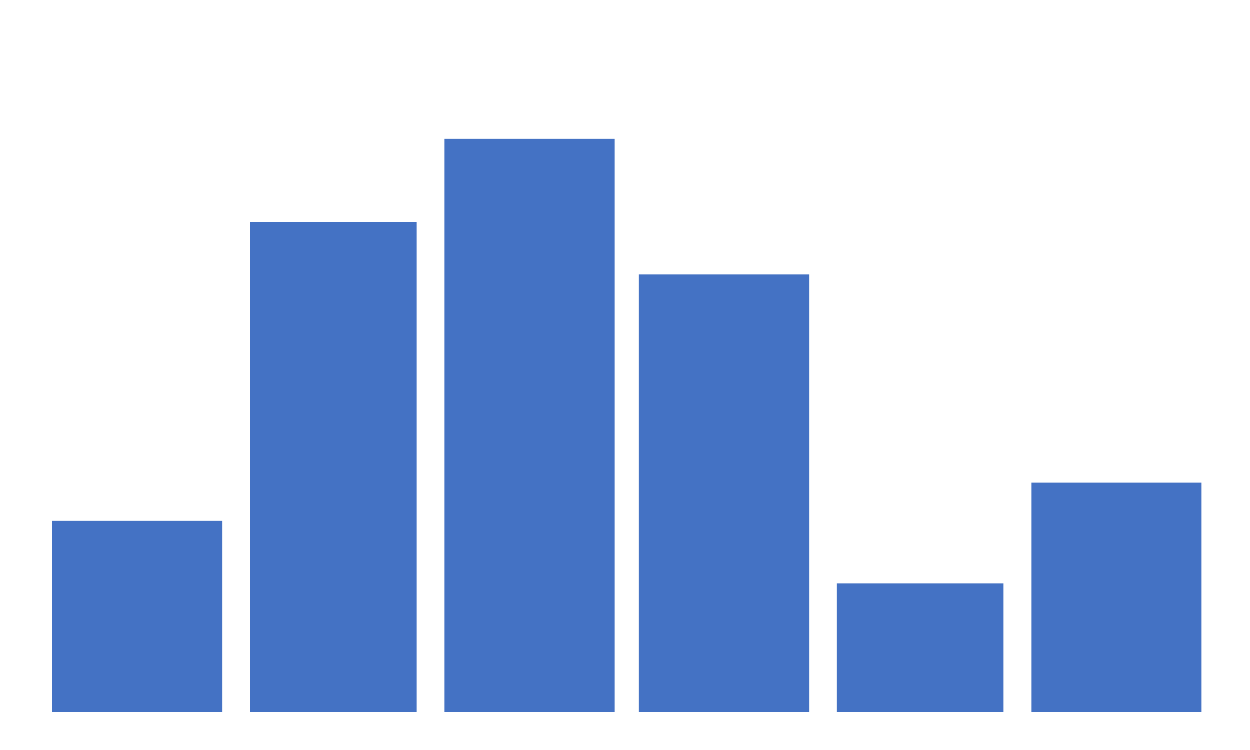}  &3.75 & .244 & 
-0.33 & 1.00 & 0.01 & 1.00 & 0.10  \\

\hline

\multicolumn{10}{l}{\textbf{T2. Characteristics of Accessibility}}\\
\hline

Accessibility is intertwined with multiple activities. & S8 & \includegraphics[width = 0.4cm, height = 0.12  cm]{Bigtable_figures/Concept_2.pdf} & 4.19 & \faWrench \ \textbf{.000} & 
\cellcolor[rgb]{ .921,  .921,  .921}-0.43  & .793 & 0.18 & .435 & 0.32  \\

Accessibility is dynamic in nature. & S9 & \includegraphics[width = 0.4cm, height = 0.12  cm]{Bigtable_figures/Concept_2.pdf} & 4.01 & \faWrench \ \textbf{.000} & 
\cellcolor[rgb]{ .921,  .921,  .921}-0.13  & .631 & 0.29 & .283 & 0.45 \\
\hline

\multicolumn{10}{l}{\textbf{T3. Work Characteristics}}\\
\hline
Accessibility development requires specific knowledge and information. & S10 & \includegraphics[width = 0.4cm, height = 0.12  cm]{Bigtable_figures/Concept_1.pdf} & 4.25 & 1.00 & 0.04 & 1.00 & 0.17 & .922 & -0.24  \\

Interaction with outside organizations is needed. & S11 & \includegraphics[width = 0.4cm, height = 0.12  cm]{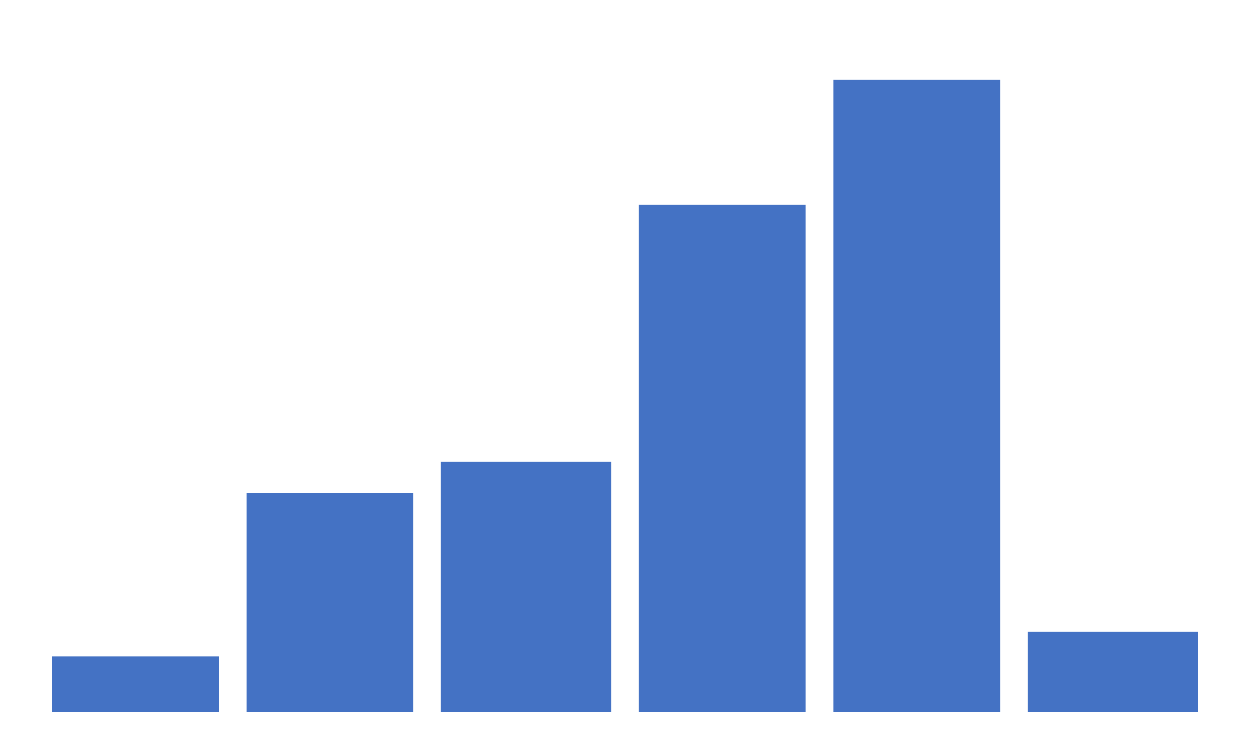} & 4.07 & .148 & 
0.18 & .280 & -0.34 & .255 &0.21 \\

Accessibility task identification is time-consuming. & S12 & \includegraphics[width = 0.4cm, height = 0.12  cm]{Bigtable_figures/Concept_3.pdf}  & 4.01 & 1.00 & 
.003 & 1.00 & 0.05 & \phone \ \textbf{.002} & \cellcolor[rgb]{ .664,  .664,  .664} 0.39  \\

\hline

\multicolumn{10}{l}{\textbf{T4. Organizational Factors }}\\
\hline

Accessibility is a marketing strategy.  & S13 & \includegraphics[width = 0.4cm, height = 0.12  cm]{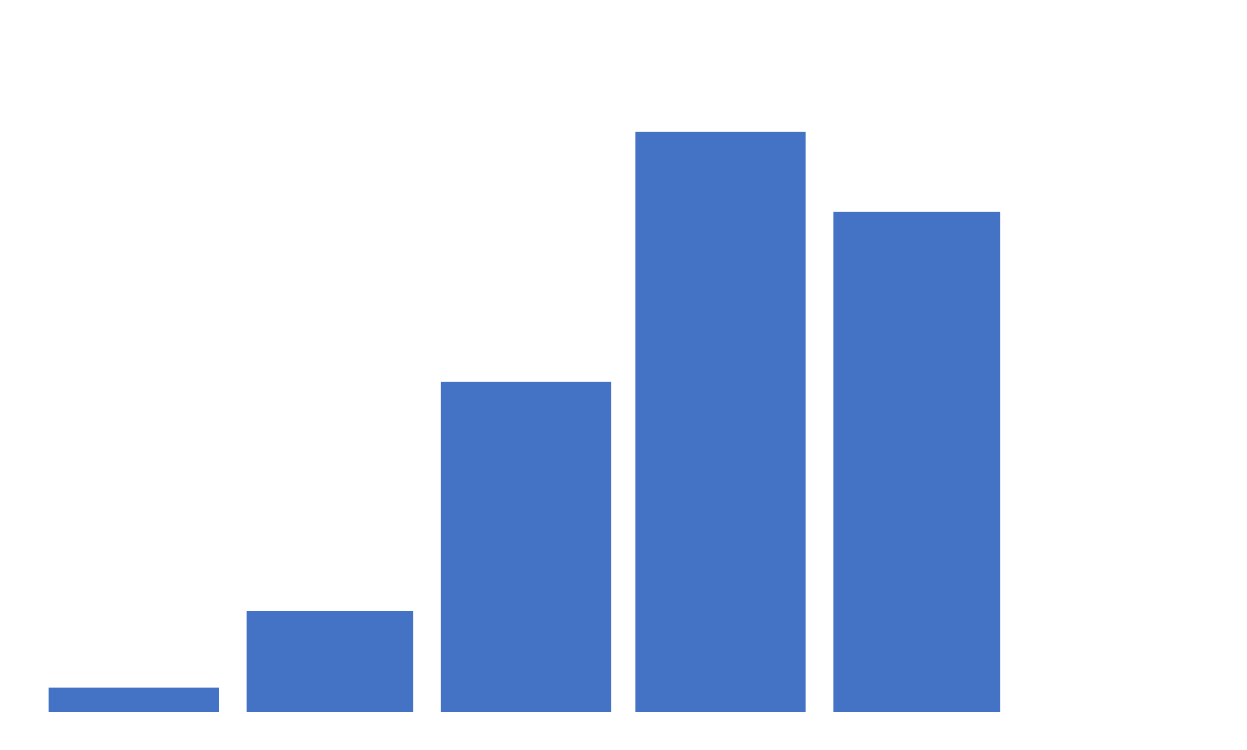} & 4.27 & .979 & 0.07 & .897 & 0.11 & .265 & 0.06 \\

Most commercial and mature projects abide by accessibility design principles. & S14 & \includegraphics[width = 0.4cm, height = 0.12  cm]{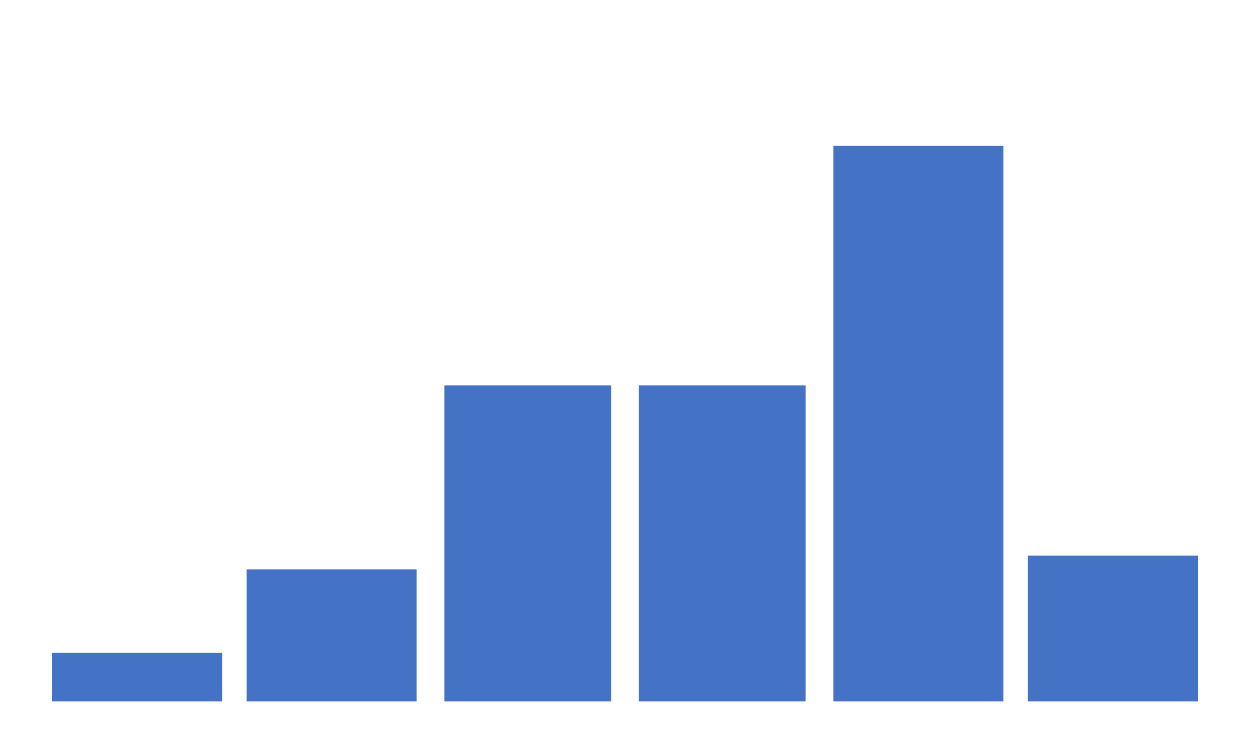} & 4.13 & .063 & -0.12
 &.894   & -0.01& .564 & 0.04  \\

For some big companies, accessibility is not optional but a key task. & S15 & \includegraphics[width = 0.4cm, height = 0.12  cm]{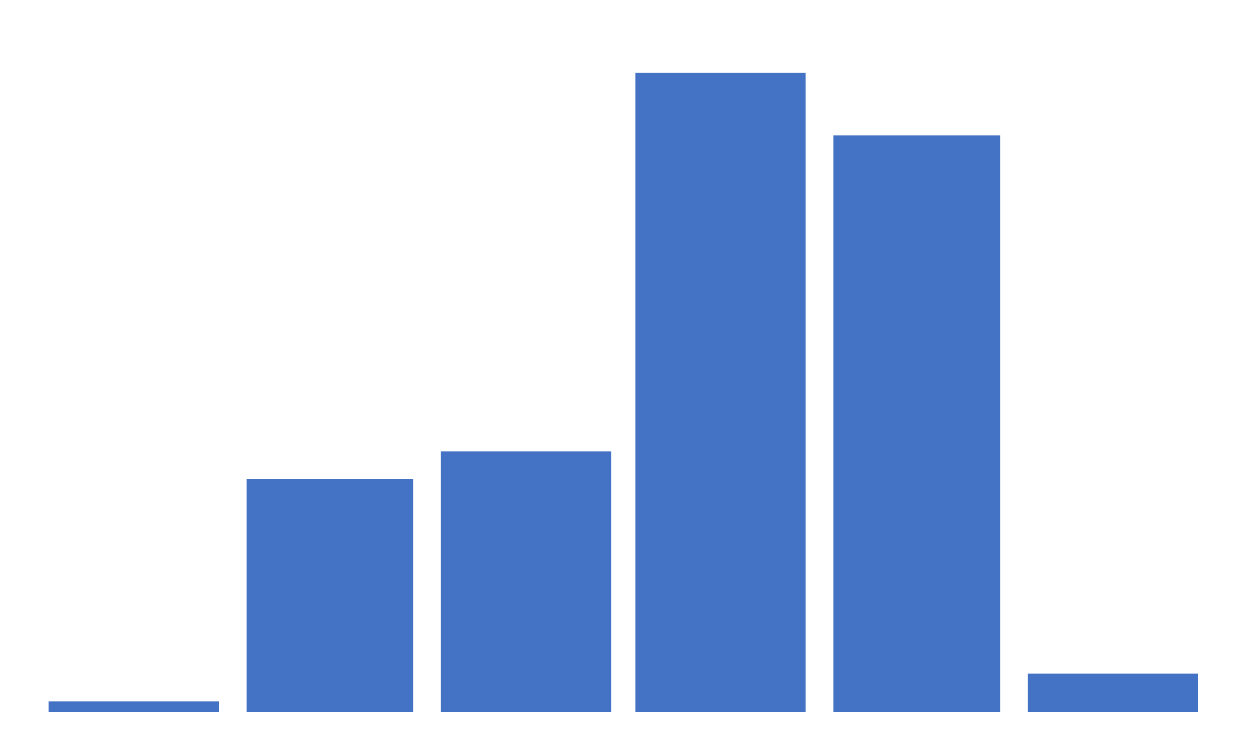}  & 4.10 & .152 & 
  0.07 & .151 & 0.11 & .854 & -0.05\\

Accessibility has lack of demand in the industry.  & S16 & \includegraphics[width = 0.4cm, height = 0.12  cm]{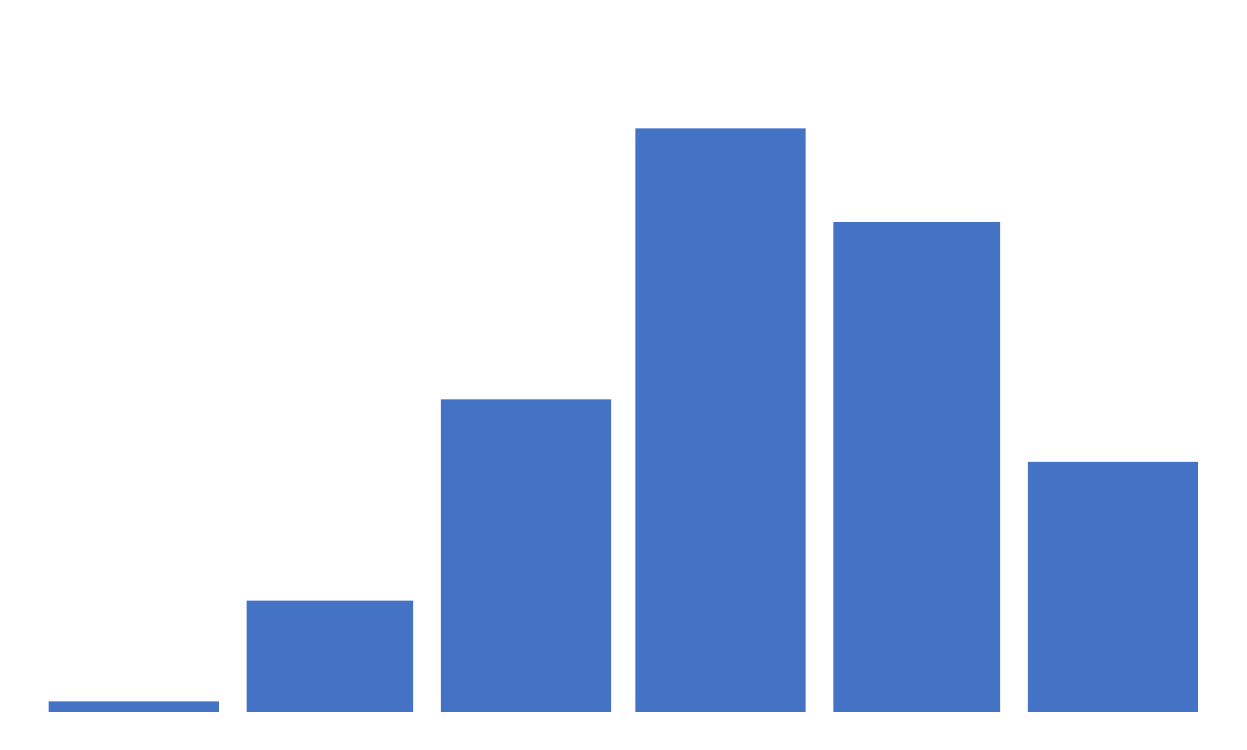} & 4.05 &  \faWrench \ \textbf{.001} &  \cellcolor[rgb]{ .664,  .664,  .664} 0.19  & \faGroup \ \textbf{.001}  & \cellcolor[rgb]{ .921,  .921,  .921} -0.10 & \phone \ \textbf{.002}  &   \cellcolor[rgb]{ .664,  .664,  .664} 0.45   \\

 Accessibility design lacks financial and organizational support. & S17 & \includegraphics[width = 0.4cm, height = 0.12  cm]{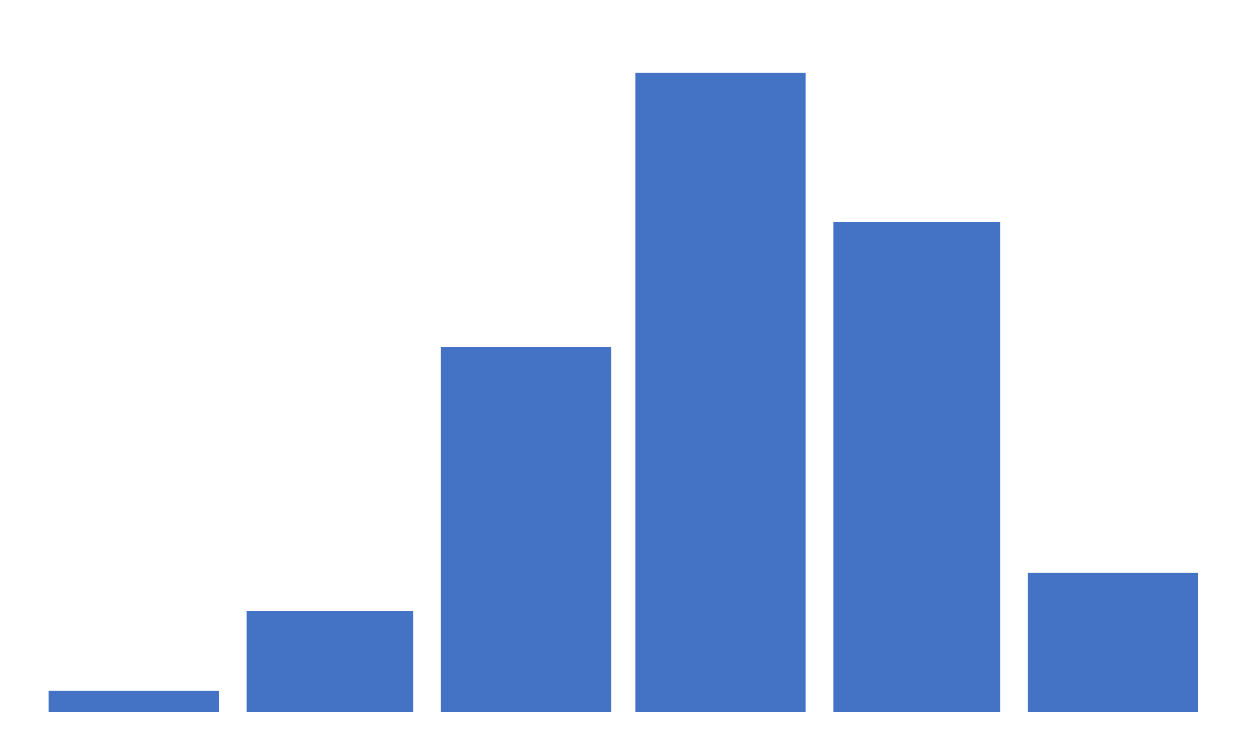} & 4.02 & 
.107 & 0.10 & .069 & 0.09 & .068 & -0.08\\

 Accessibility is context-dependent. & S18 & \includegraphics[width = 0.4cm, height = 0.12  cm]{Bigtable_figures/Concept_3.pdf}  & 4.01 & 1.00 & 
0.09 & .806 & 0.22 & .079 & 0.38  \\

Small companies (teams) do not consider accessibility design. & S19 & \includegraphics[width = 0.4cm, height = 0.12  cm]{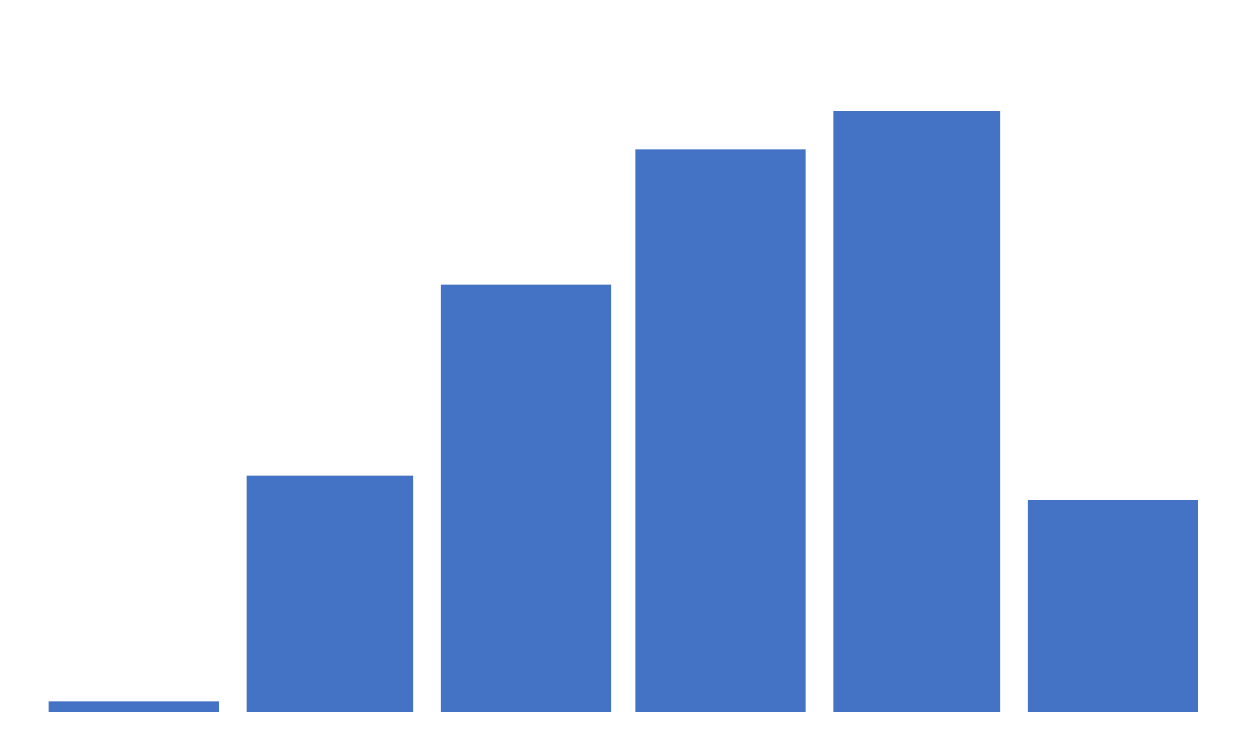}  & 3.82 & 
.063 & 0.15 & .995 & 0.04 & .166 & 0.15 \\

There are limited relevant guidelines for accessibility design. & S20 & \includegraphics[width = 0.4cm, height = 0.12  cm]{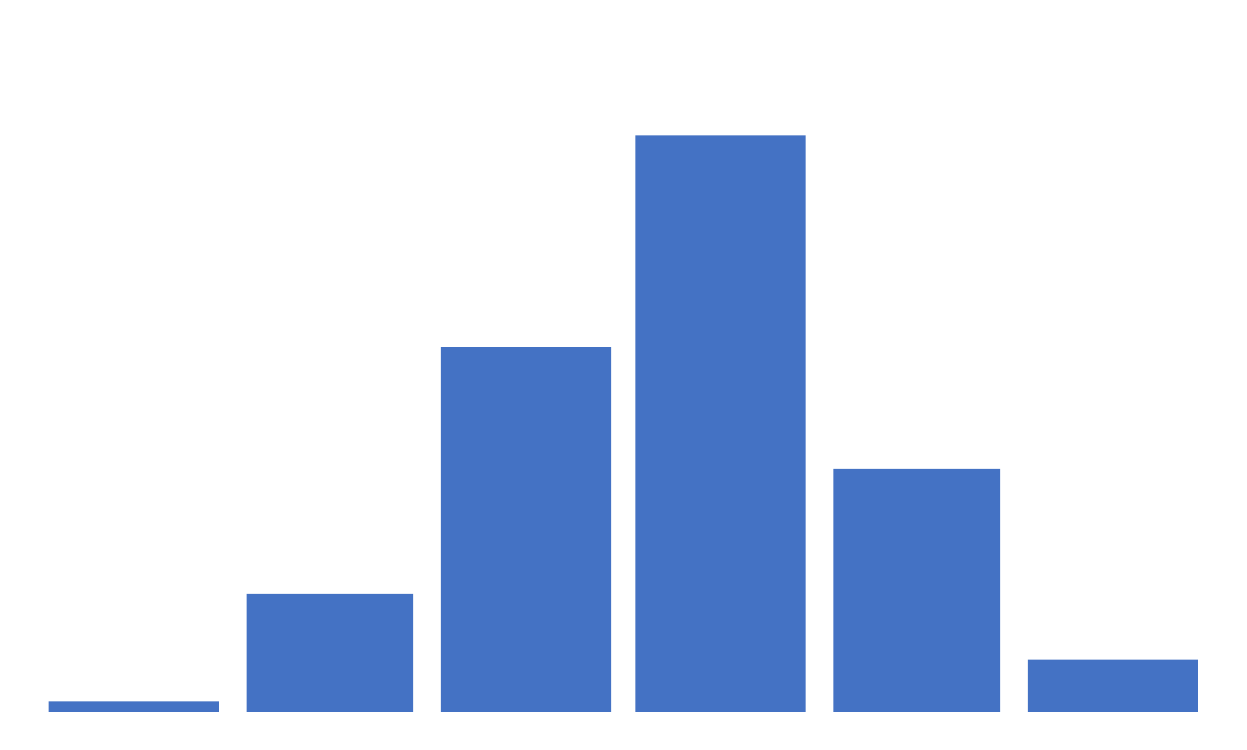}  & 3.70 & .947 & 
0.02 & .282 & -0.16 & .108 & 0.31 \\

Supporting accessibility minimizes legal risks.   & S21 & \includegraphics[width = 0.4cm, height = 0.12  cm]{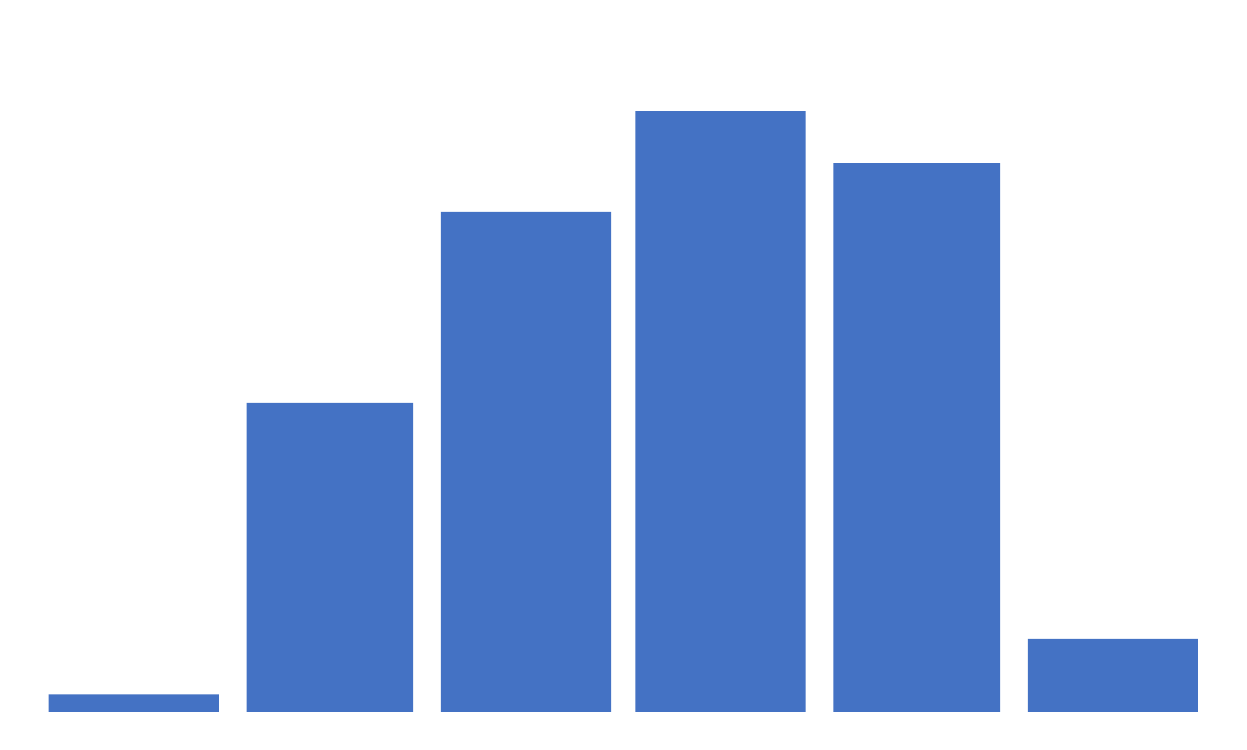}  & 3.38 & .056 & 
0.13 & .320 & -.019 & .271 & 0.18  \\

\hline

\multicolumn{10}{l}{\textbf{T5. Accessibility Requirement}}\\
\hline

 Accessibility requirements are an expensive addition for the system.  & S22 & \includegraphics[width = 0.4cm, height = 0.12  cm]{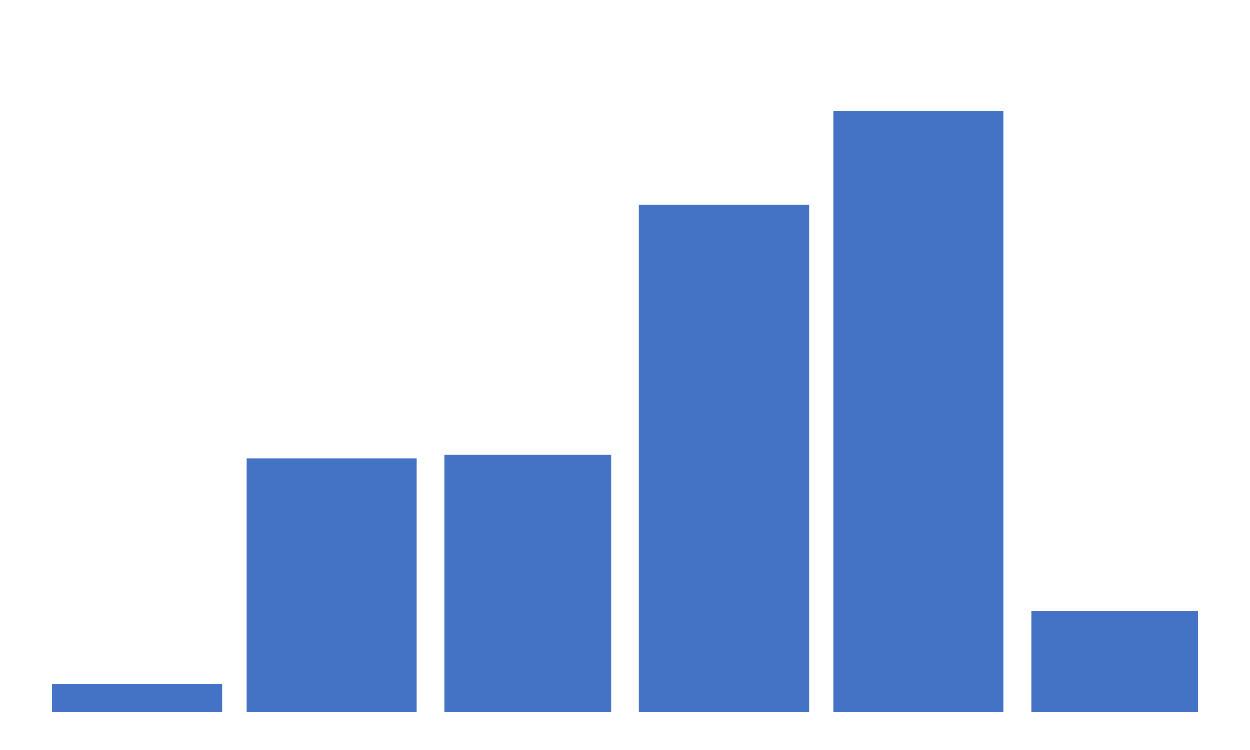} & 4.10 & .233 & .090 & 0.10 & .251 & .057 & 0.27  \\

Accessibility requirements are more focused on FR design. & S23 & \includegraphics[width = 0.4cm, height = 0.12  cm]{Bigtable_figures/Requirement_2.pdf} & 4.05 & .121 & 
-0.14 & 1.00 & -0.65 & 1.00 & 0.71 \\

Accessibility requirements are not clearly documented. & S24 & \includegraphics[width = 0.4cm, height = 0.12  cm]{Bigtable_figures/Requirement_2.pdf} & 4.07 & .148 & 
0.18 & 0.280 & -0.34 & .255 &0.21 \\

No resources from companies for accessibility requirements elicitation. & S25 & \includegraphics[width = 0.4cm, height = 0.12  cm]{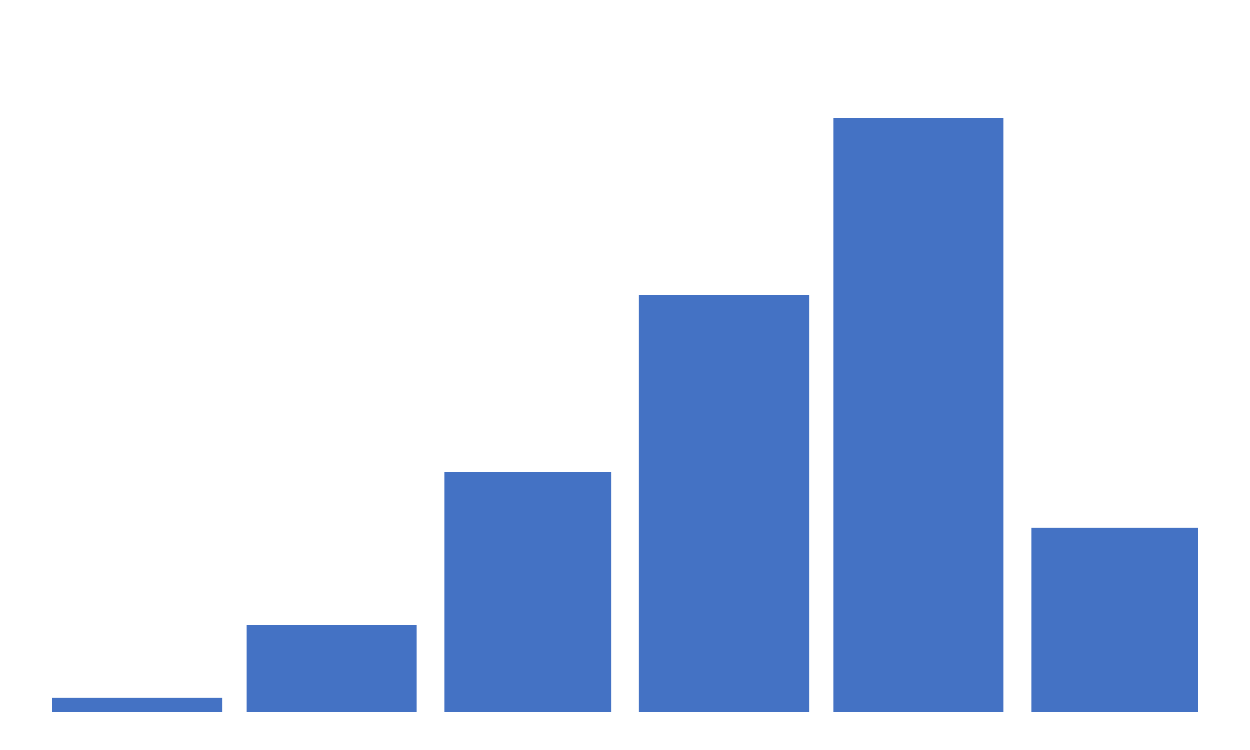} & 4.03 & .186 & 
-0.34 & \faGroup \ \textbf{.004} & \cellcolor[rgb]{ .921,  .921,  .921} -0.23 & 1.00 & 0.04 \\

Not a core requirement for the project. & S26 & \includegraphics[width = 0.4cm, height = 0.12  cm]{Bigtable_figures/Requirement_3.pdf} & 4.00 & \faWrench \ \textbf{.002} & 
 \cellcolor[rgb]{ .921,  .921,  .921} -0.14 & 0.63 & 0.15 & .241 & 0.22 \\

 ML and AL can be applied to collect requirements regarding accessibility. & S27 & \includegraphics[width = 0.4cm, height = 0.12  cm]{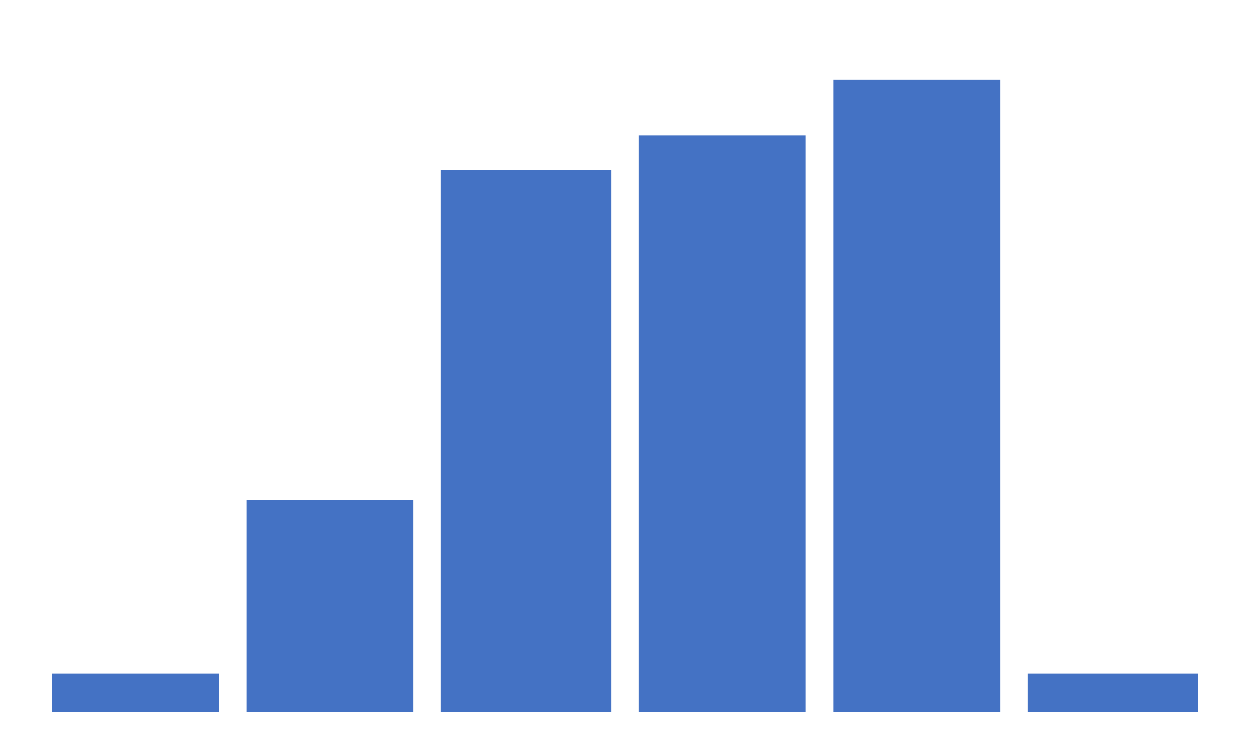} & 3.77 & 
1.00 & 0.04 & .076 & 0.26 &1.00 &  -0.25  \\

 Difficult to understand the technologies behind the requirements. & S28 & \includegraphics[width = 0.4cm, height = 0.12  cm]{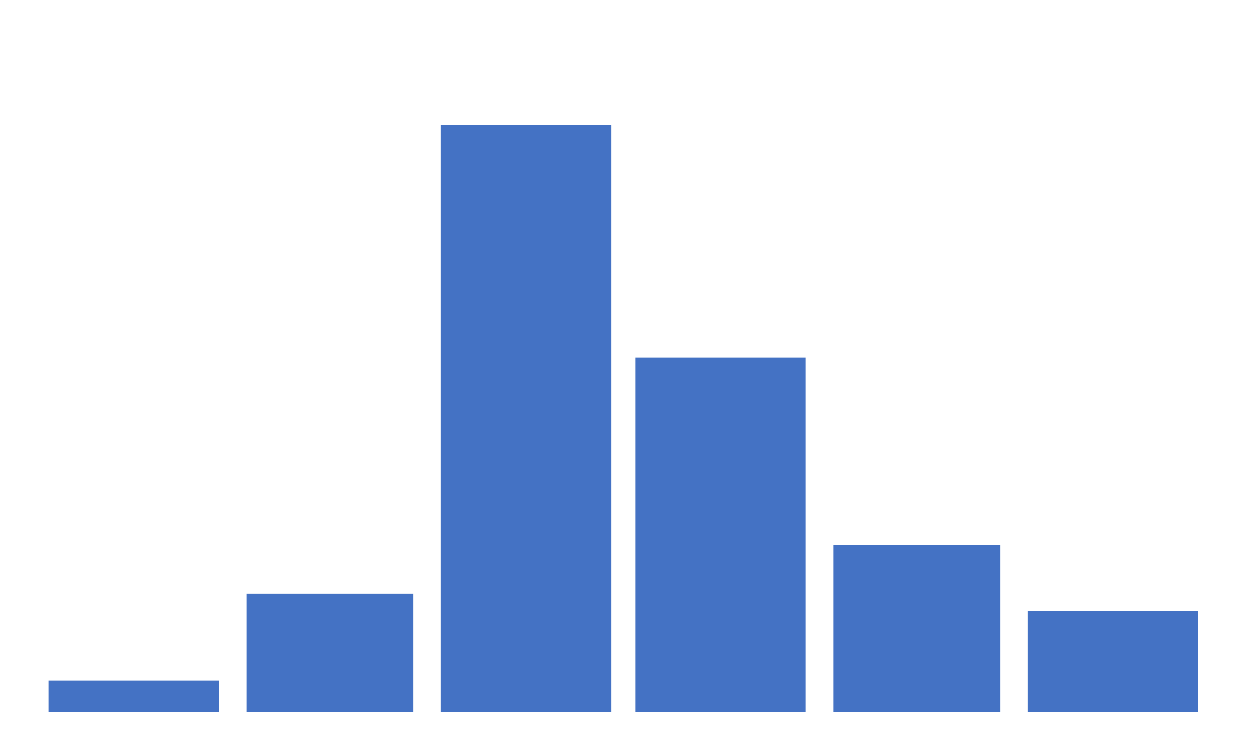} & 3.40 & .186 & 
-.034 & .394 & -0.30 & .435 & 0.32  \\

\hline

\multicolumn{10}{l}{\textbf{T6. Accessibility Design}}\\
\hline

 Provide multiple views to address trade-offs between different types of user groups. & S29 & \includegraphics[width = 0.4cm, height = 0.12  cm]{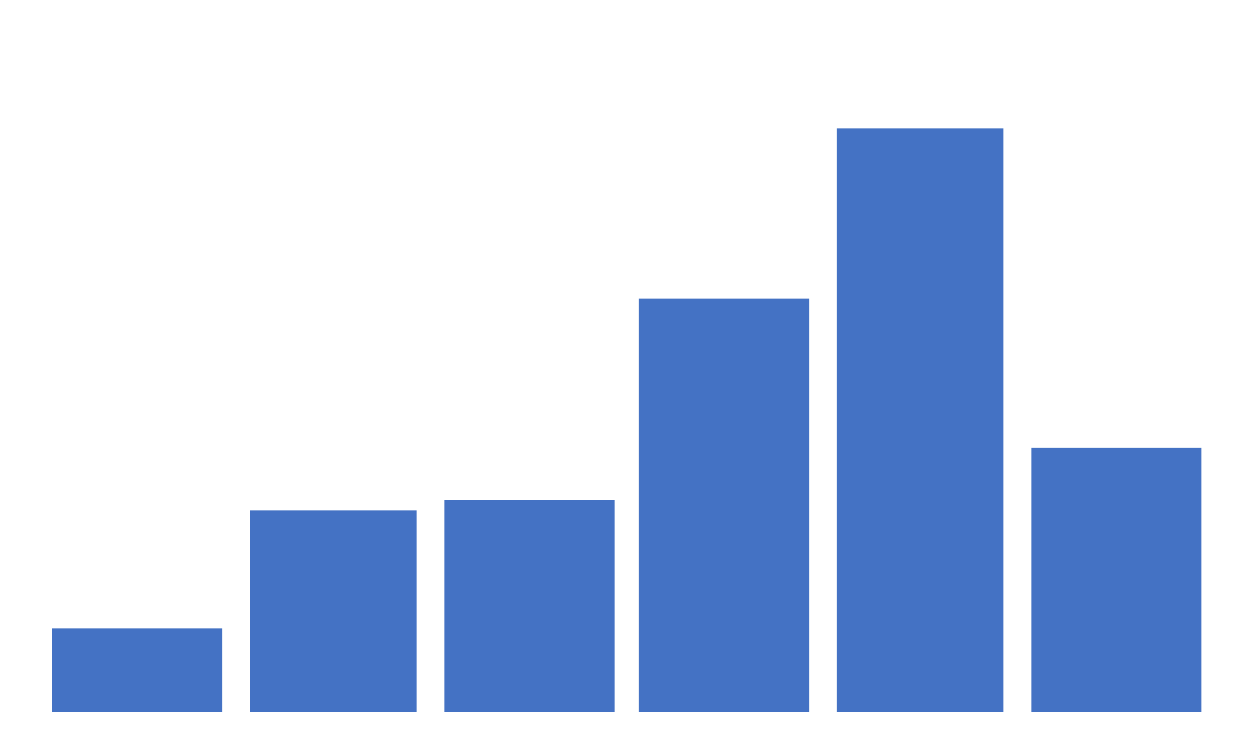} & 4.01 & 1.00 & 0.06 & 1.00 & 0.21 & 1.00 & -0.03  \\

Front-end design is often the major focus to make sure projects are accessible. & S30 & \includegraphics[width = 0.4cm, height = 0.12  cm]{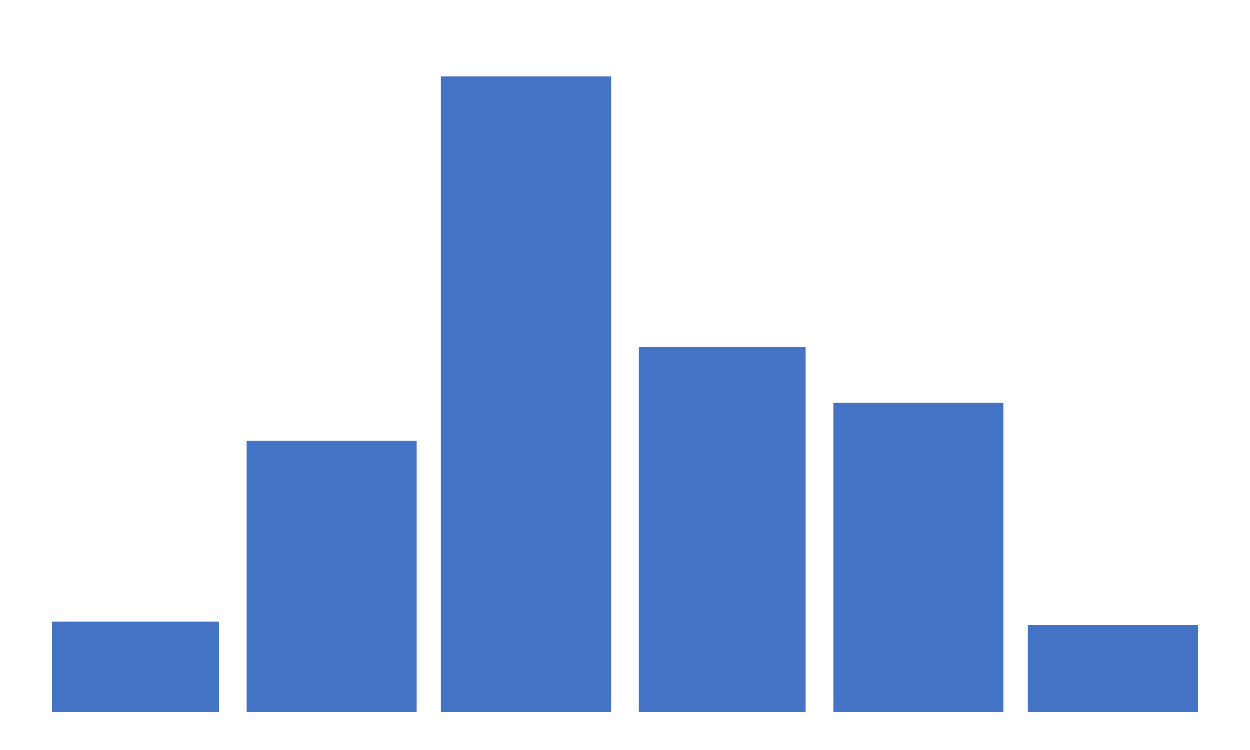} & 3.98 & \faWrench \ \textbf{.003} & 
\cellcolor[rgb]{ .664,  .664,  .664}  0.11  & .077 & 0.07 & .540 & 0.04 \\

ML and AI technologies could be applied to help accessibility design. & S31 & \includegraphics[width = 0.4cm, height = 0.12  cm]{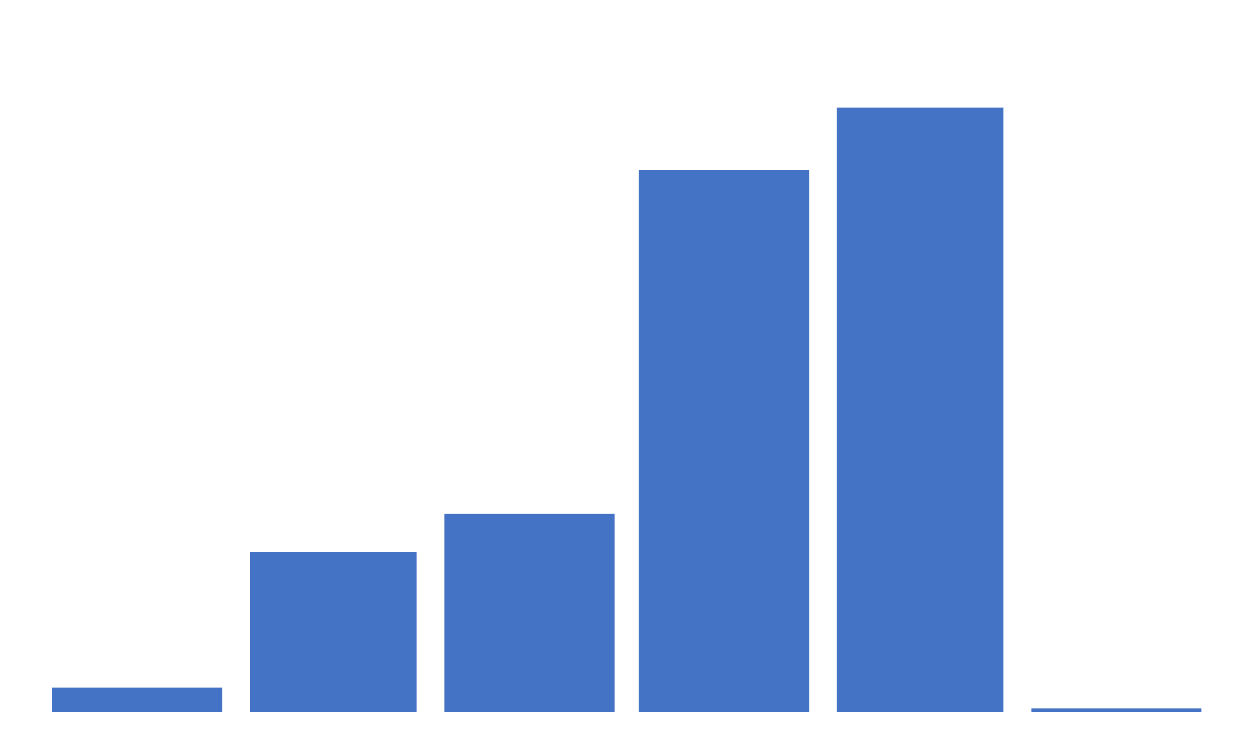} & 3.82 &  
.235 & -0.02 & 1.00 & -0.12 & .280 & 0.29\\

Front-end design is often the major responsibility to make sure projects are accessible. & S32 & \includegraphics[width = 0.4cm, height = 0.12  cm]{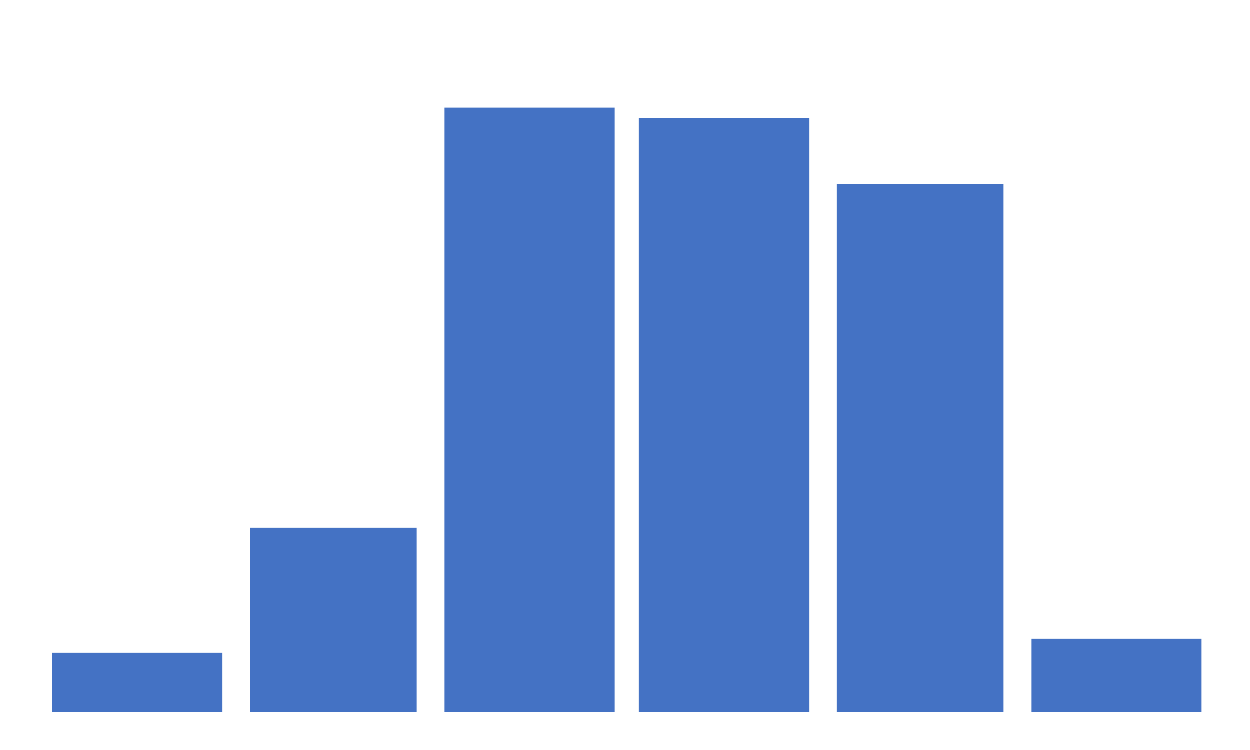} & 3.76 &  
.999 & -0.17 & .251 & 0.21 & .079 & 0.48 \\

 Detailed design is time-consuming and conducted in an iterative way. & S33 & \includegraphics[width = 0.4cm, height = 0.12  cm]{Bigtable_figures/Concept_2.pdf} & 3.71 & .138 & 
.094 & \faGroup \ \textbf{.000} & \cellcolor[rgb]{ .921,  .921,  .921} -0.34 & .266 & 0.07 \\

 Hard to make accessibility design decisions (like design patterns or tactics). & S34 & \includegraphics[width = 0.4cm, height = 0.12  cm]{Bigtable_figures/Concept_2.pdf} & 3.70 & 0.02 & 
-0.36 & .388 & 0.26 & \phone \ \textbf{.001} & \cellcolor[rgb]{ .921,  .921,  .921}-0.33 \\

Some accessibility standards are out of date. & S35 & \includegraphics[width = 0.4cm, height = 0.12  cm]{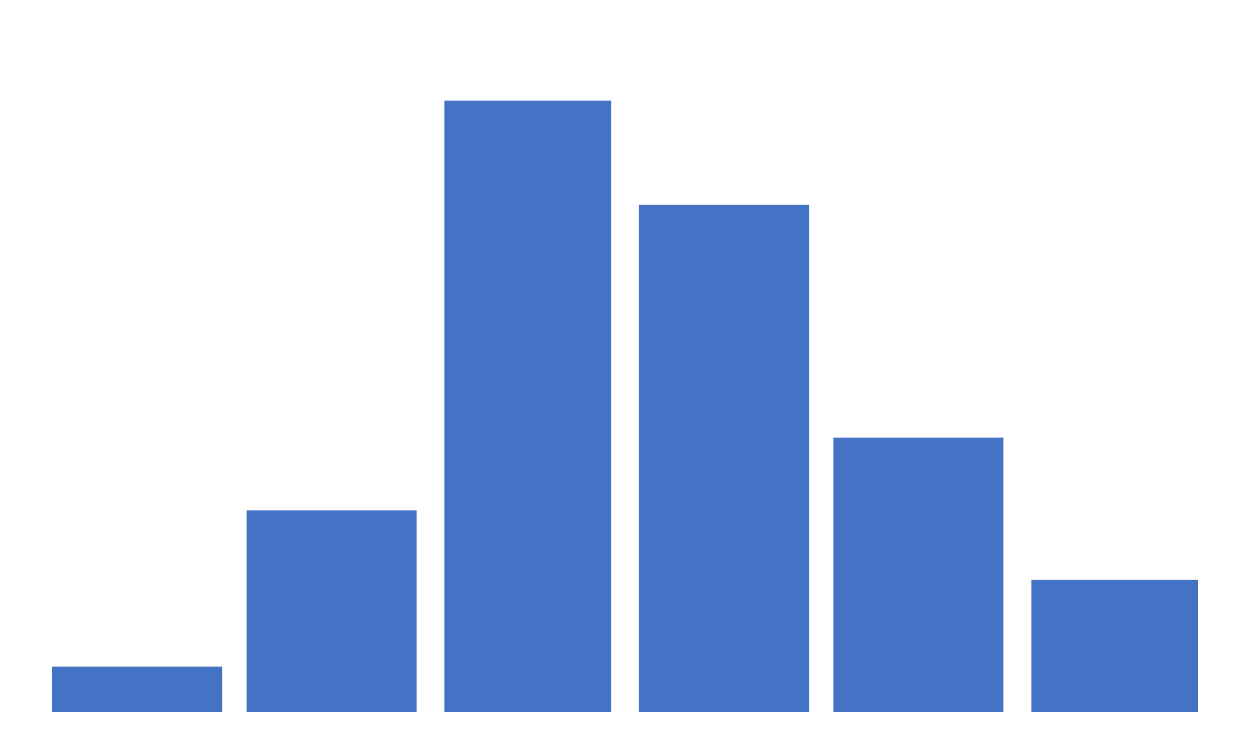} & 3.65 & 
.897 & -0.23 & .674 & -0.19 & 1.00 & 0.17 \\

\hline

\multicolumn{10}{l}{\textbf{T7. Accessibility Testing}}\\

\hline

 A long list for accessibility testing (FRs and NFRs).  & S36 & \includegraphics[width = 0.4cm, height = 0.12  cm]{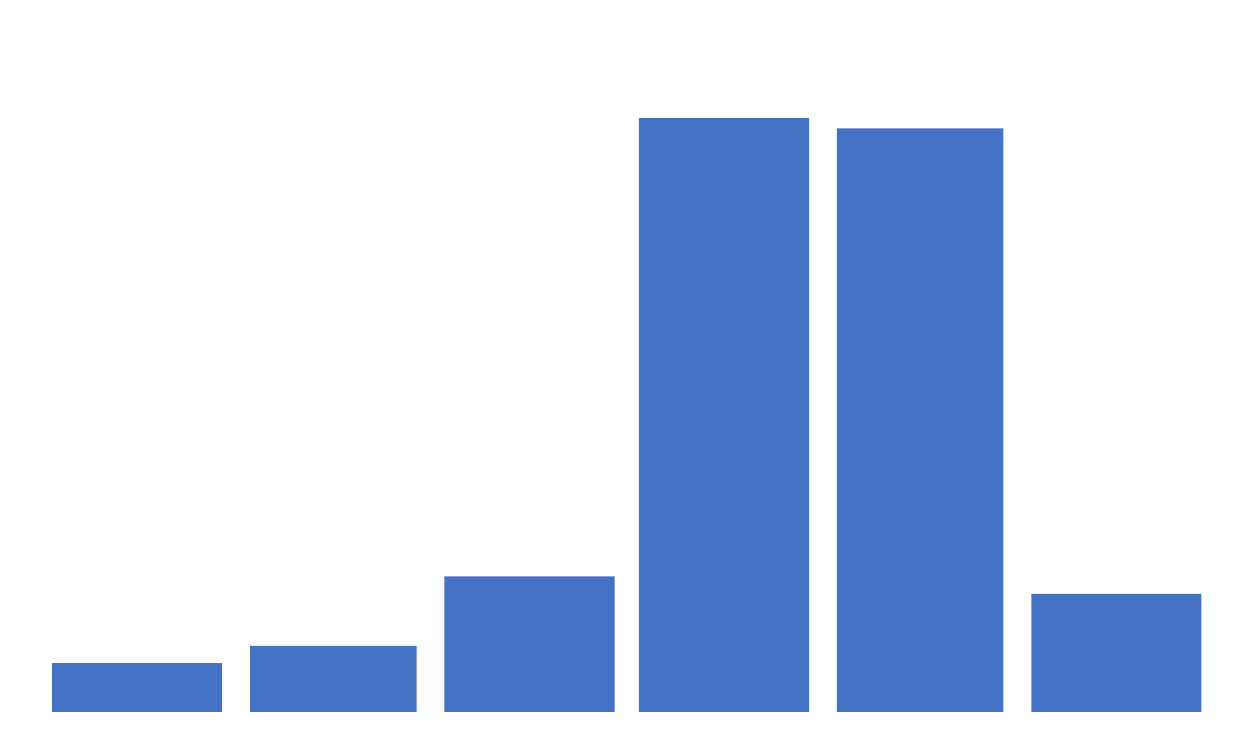} & 4.25 & .387 & 0.28 & 1.00 & -0.45 & .281 & -0.28  \\

 It is hard to engage with end-users (get feedback). & S37 & \includegraphics[width = 0.4cm, height = 0.12  cm]{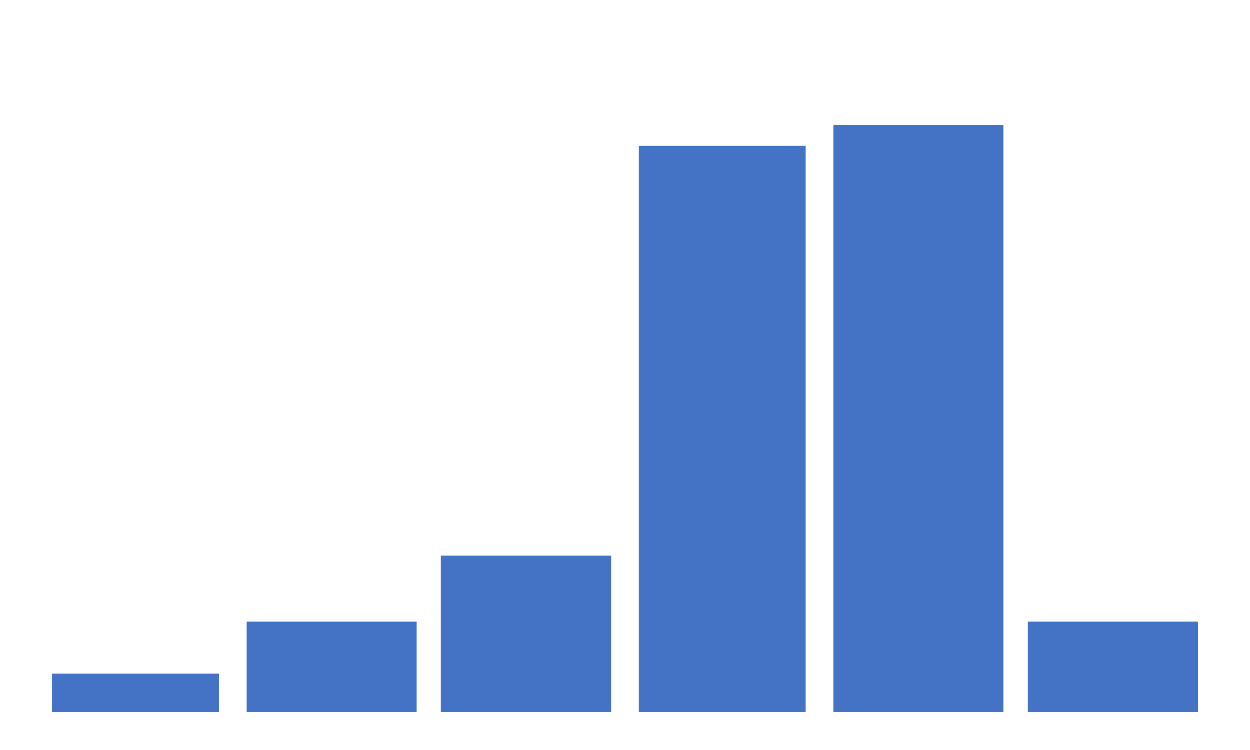} & 4.19 & 1.00 & 
0.06 & 1.00 & 0.04 & .079 & 0.28 \\

 Automatic testing tools are useful.  & S38 & \includegraphics[width = 0.4cm, height = 0.12  cm]{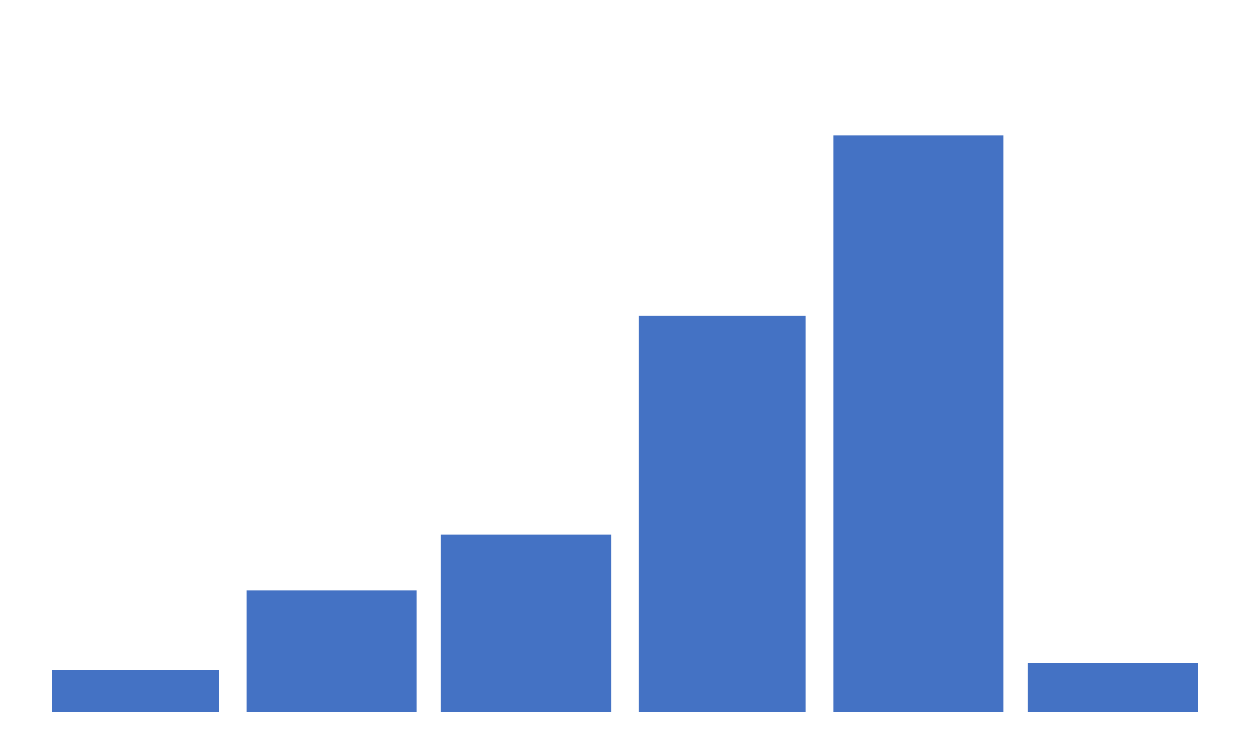} & 4.19 & .035 & 
-034 & .871 & 0.23 &\phone \ \textbf{ .000} & \cellcolor[rgb]{ .921,  .921,  .921}-0.15  \\

\hline

\multicolumn{10}{l}{\textbf{T8. Accessibility Evaluation}}\\
\hline

 No single tool can determine if a site or project meets accessibility guidelines. & S39 & \includegraphics[width = 0.4cm, height = 0.12  cm]{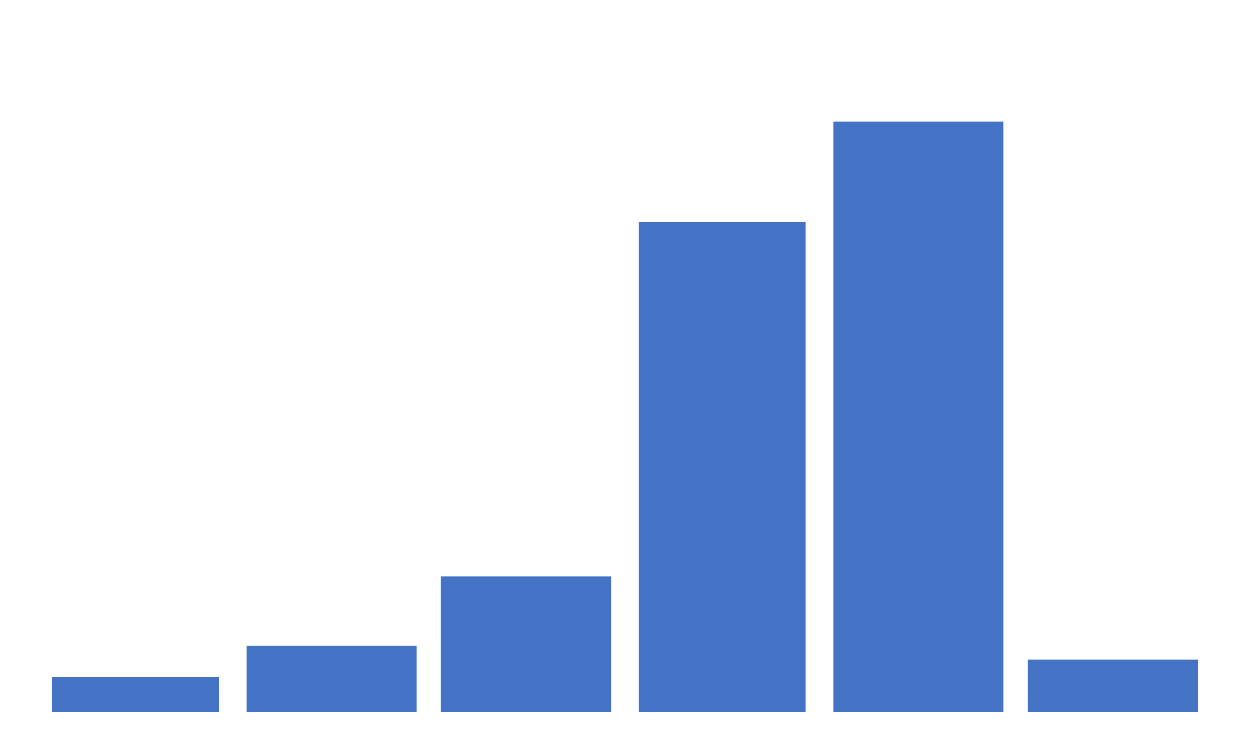} & 4.17 & 1.00& -0.14 & 1.00 & 0.17 & 1.00 & 0.10  \\

Human evaluation is always required. & S40 & \includegraphics[width = 0.4cm, height = 0.12  cm]{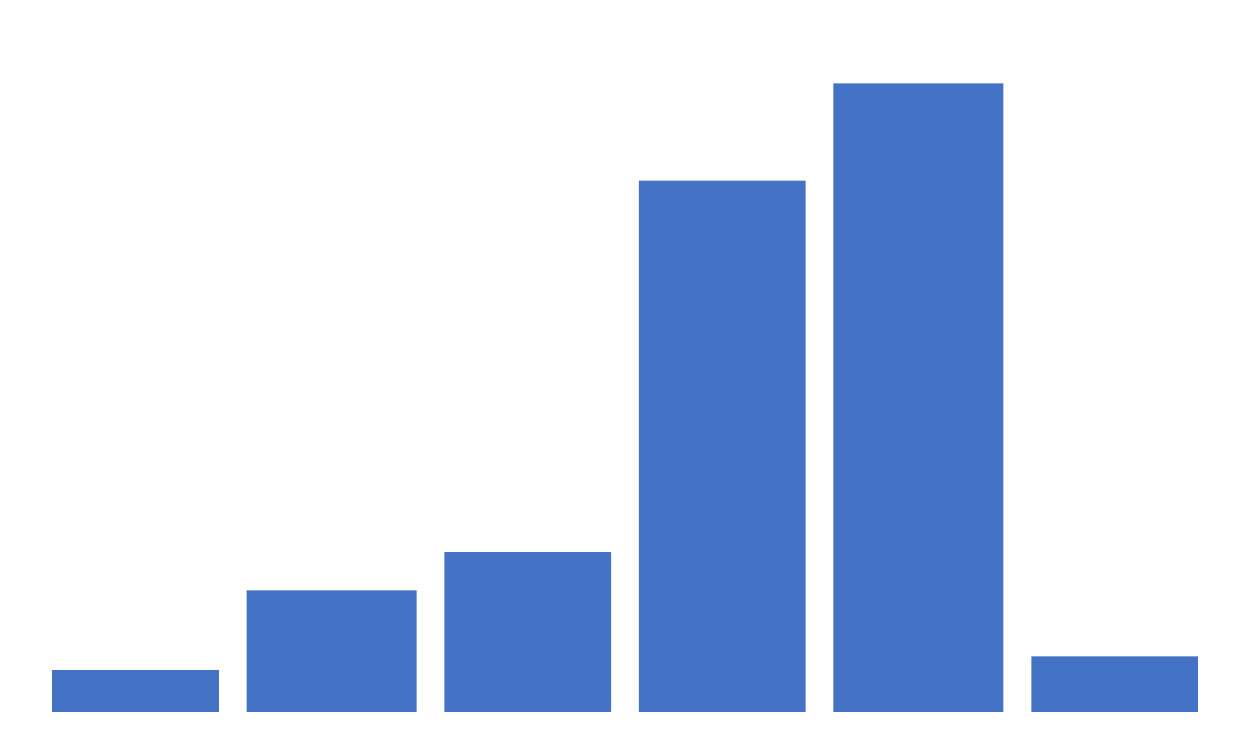} & 4.10 & .266 & 
0.13 & .498 & 0.11 & .719 & -0.02 \\

 A lot of extra effort is needed for accessibility evaluation. & S41 & \includegraphics[width = 0.4cm, height = 0.12  cm]{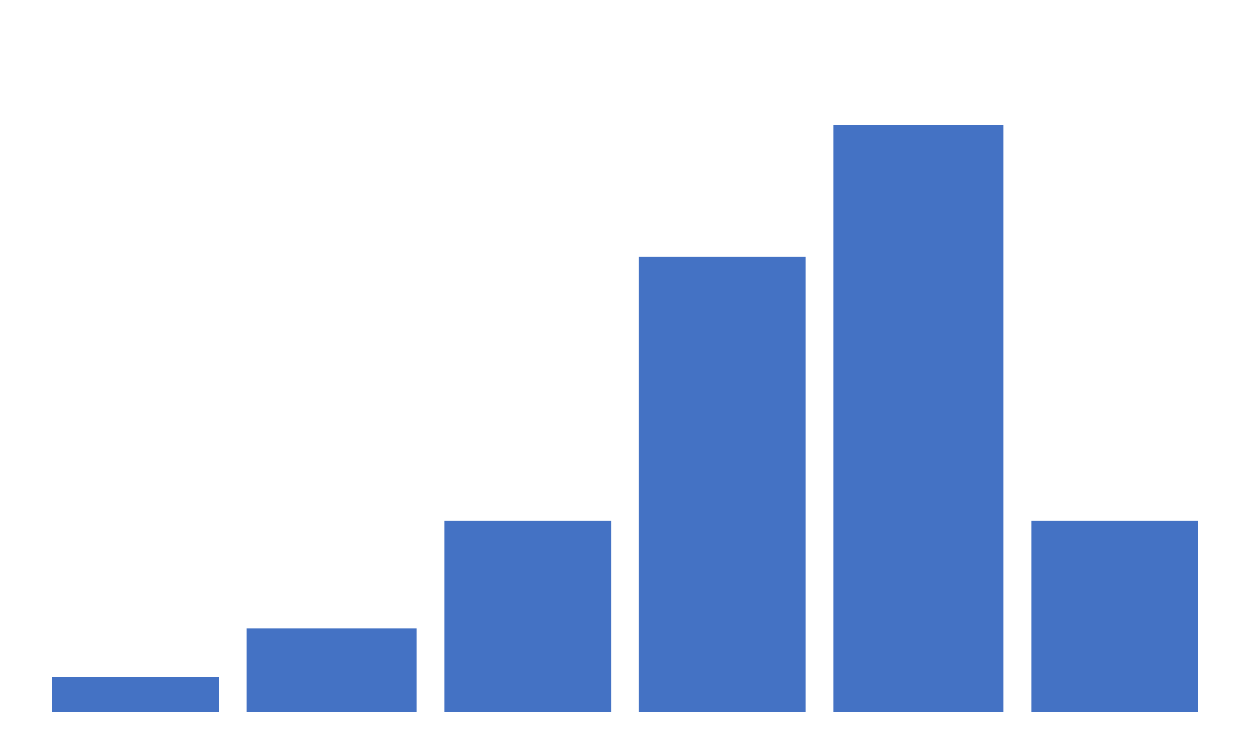} & 4.10 
& .719 & 0.06 & .665 & 0.08 & 0.65 & -0.21 \\

 It is difficult to get feedback from end-users for accessibility evaluation. & S42 & \includegraphics[width = 0.4cm, height = 0.12  cm]{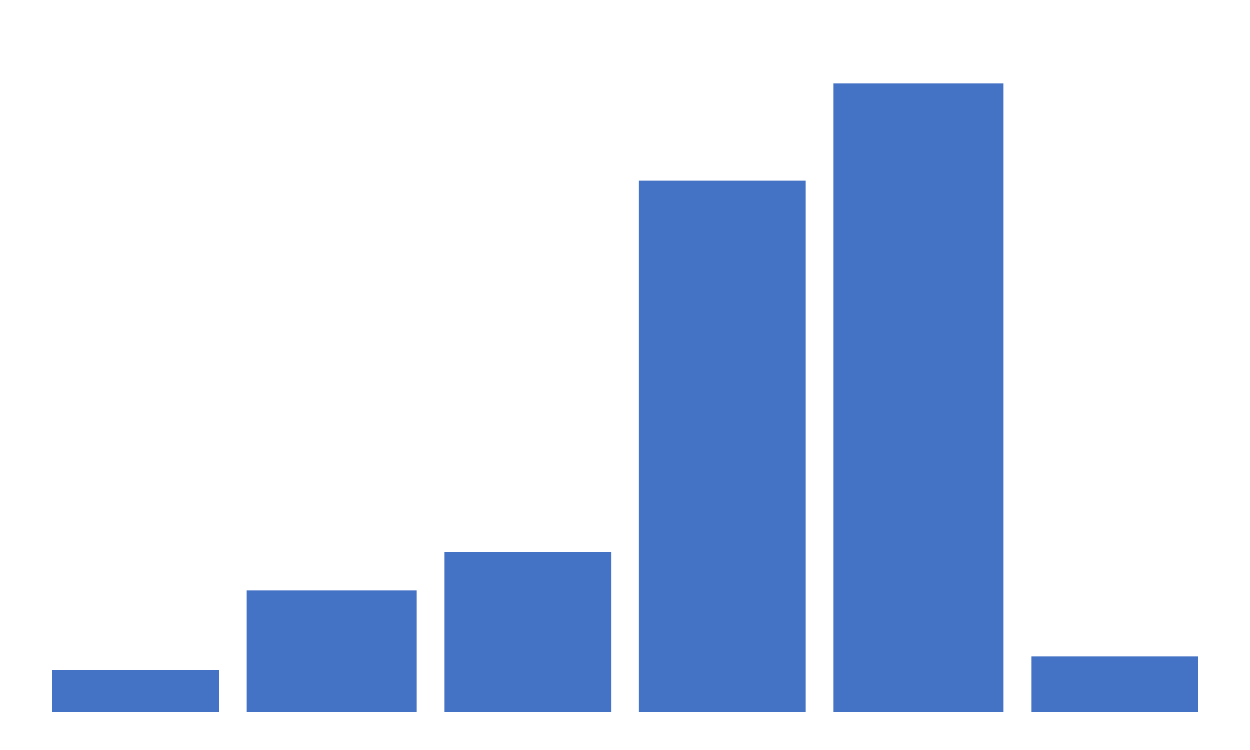} & 4.19 & 
.320 & -0.19 & .275 & 0.19 & .182 & -0.25 \\

A more comprehensive evaluation for accessibility is necessary. & S43 & \includegraphics[width = 0.4cm, height = 0.12  cm]{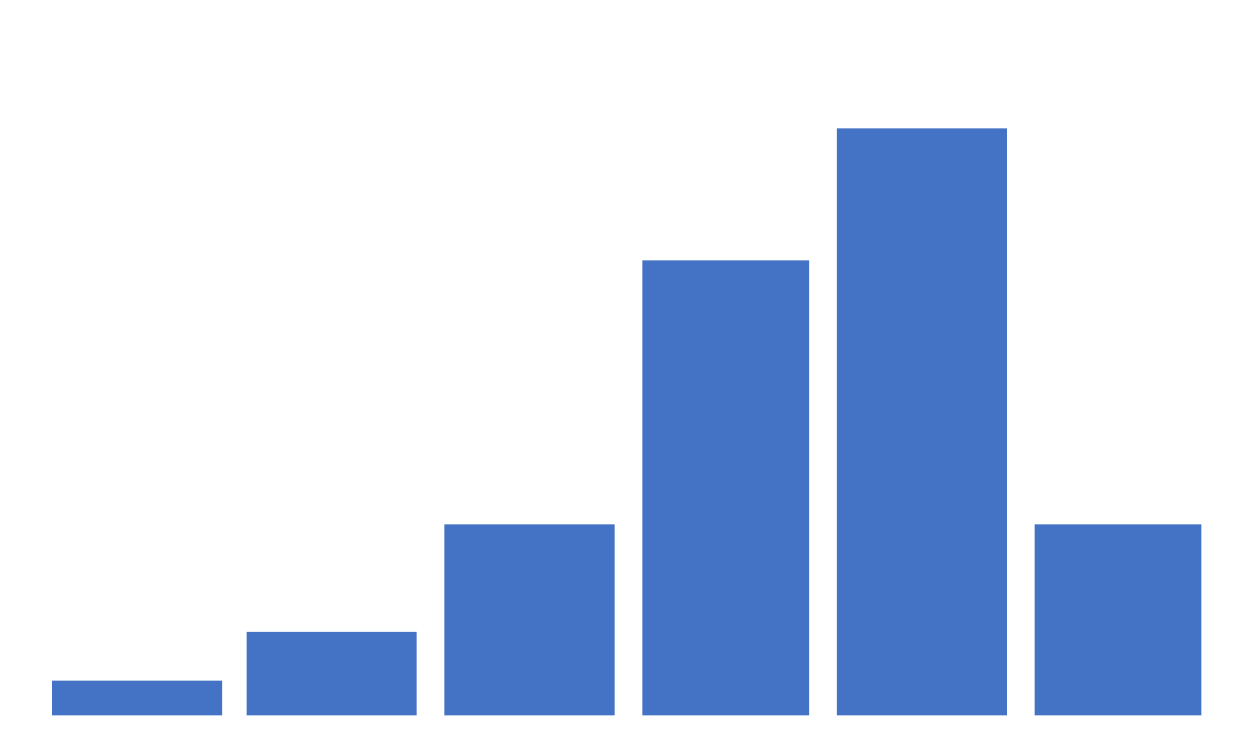} & 4.01 & 
.797 & 0.05 & \faGroup \ \textbf{.001} & \cellcolor[rgb]{ .664,  .664,  .664} 0.14 & .546 & 0.04 \\

 Standards (e.g., WCAG) are helpful for accessibility evaluation. & S44 & \includegraphics[width = 0.4cm, height = 0.12  cm]{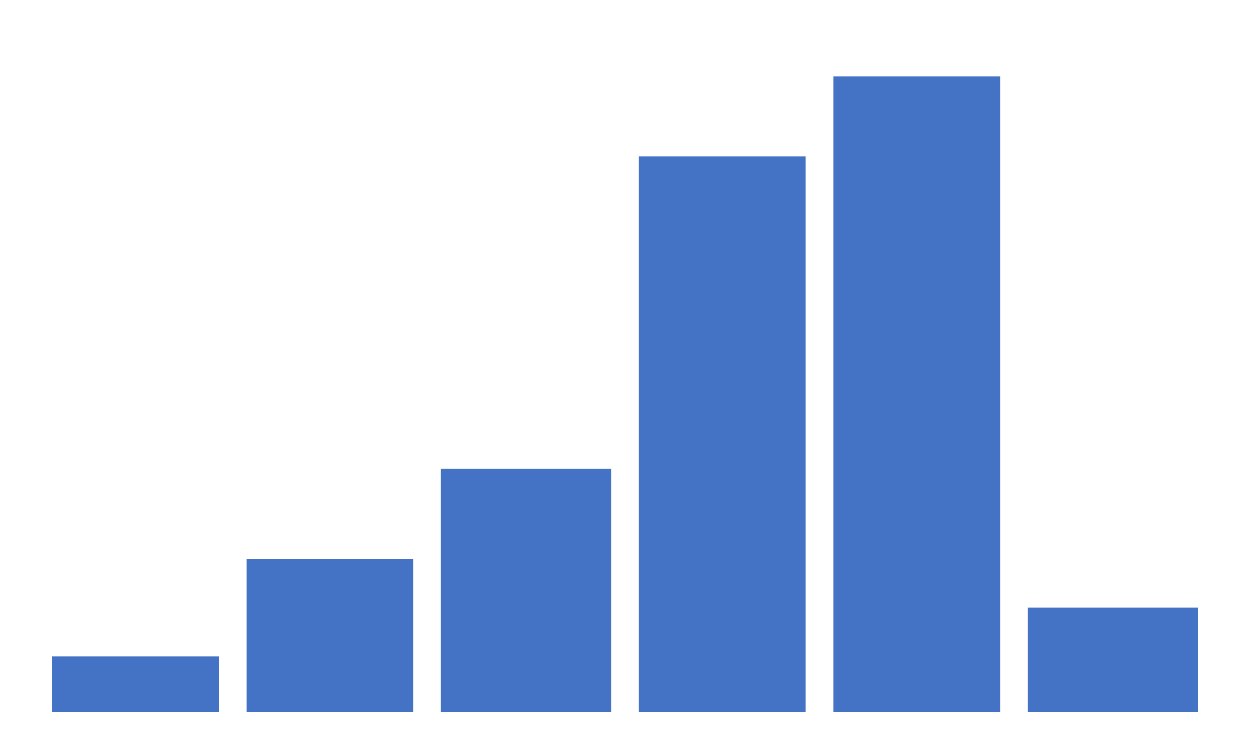} & 3.99 & 1.00 & 
0.06 & 1.00 & 0.40 & \phone \ \textbf{.000 }&  \cellcolor[rgb]{ .664,  .664,  .664}0.13\\

\hline
 
\end{tabular}
\end{adjustbox}
\end{table}

\end{landscape}

\subsubsection{Understanding accessibility}
\label{SubsecRQ1_ Understanding}

We explored how practitioners understand accessibility, for example, what motivates them to address accessibility issues for end users, and what deters them from incorporating accessibility into their projects. We analyzed their opinions on the importance and characteristics of accessibility development and design in practice. 

\textbf{General consideration of accessibility}. Regarding the general considerations of accessibility, we summarized seven statements in Table \ref{Table2_Statements} from the interviews. We ranked the statements by scores, which was derived from the online survey results. Our interviewees' opinions on understanding accessibility are consistent with ISO 9241 \cite{jokela2003standard}, which emphasizes that the goals of accessibility design are maximizing the number of users and striving to increase the level of usability. Accessibility can be provided by a combination of both software and hardware controlled by projects. Assistive technologies typically provide specialized input and output capabilities not provided by a typical software and hardware system. Software examples include on-screen keyboards that replace physical keyboards and voice and gesture recognition to replace keyboards and mice. Hardware examples include head-mounted pointing devices instead of mice and Braille output devices instead of a video displays [\textbf{S5}], and 25$\%$  and 14$\%$ respondents agree and strongly agree with this statement. However, eight interviewees stated that accessibility now needs to be shifted targeting the \textbf{general} end-users and is important for every project [\textbf{S2}][\phone \ \textbf{S4}] (scores of the both statements are over 4.00). They also claimed that accessibility could be integrated into current projects as a competitive functionality [\faWrench \ \textbf{S1}][\phone \ \textbf{S6}]. Around 27$\%$ (i.e., 174 out of 365) and 28$\%$ (i.e., 102 out of 365) of our online survey respondents agree and strongly agree with \textbf{S1}. Five interviewees mentioned that even though accessibility is a key task and timing consideration, in current projects, it is not well designed, especially in relatively small-sized teams and companies and less developed countries [\faWrench \ \textbf{S7}]. Around 21$\%$ (i.e., 78 out of 365) and 8$\%$ (i.e., 29 out of 365) of respondents agree and strongly agree with this statement. 

 \faThumbsODown\  "\textit{Not everyone is familiar with accessibility and \textbf{not everyone understands why it is important}. While the principles of accessibility become known by more people every day, the concept is not on everyone's radar just yet}."  [\textit{I1- Programmer}]

 \faThumbsOUp  \ "\textit{\textbf{Accessibility should be becoming competitive functionalities} that in our development, as it now is not only for disabilities. For example, voice assistant (voice search functionalities) of Apps that we have noticed is widely used}." [\textit{I5-Designer}]
 
 \faThumbsODown\ "\textit{Accessibility is \textbf{poor addressed} by current products. We should have accessibility design elements for marketing, however, I was wondering weather actually it is used by people}."  [\textit{I2-Programmer}]
 
    \faThumbsODown\
"\textit{Current companies,\textbf{ especially small size companies}, their products address accessibility focusing on a light-weight interface design way, for example, navigation and visual hierarchy}." [\textit{I10-Designer and Manager}]

\textbf{Characteristics of accessibility development and design in practice}. Based on the results of our interviews, we summarized two main characteristics of accessibility in practice.

\textit{1}. \textit{Intertwined and iterative with certain types of activities}. Accessibility should not be an independent activity in normal software development but intertwined with many software artifacts and activities. 72$\%$ of the survey respondents agree and strongly agree that accessibility should be intertwined with various development activities, such as from the beginning of the software requirement elicitation to software testing. In addition, it could require several iterations for sufficient software accessibility development and design to be achieved in practice. Considering the difficulties inherent in developing accessible software projects and investigating how accessibility fits into the development life cycle is vital and also the goal of this paper [\faWrench \ \textbf{S8}]. We expanded this statement in RQ2 and provided detailed results in Section \ref{results of RQ2}.

 \faThumbsODown\ "\textit{Issues on accessibility could creep in \textbf{all through the life cycle }of a project. The costlier accessibility issues in fact come after project completion. Countries with robust legislation and accessibility, organizations could find themselves in the midst of expensive lawsuits}." [\textit{I10-Designer and Manager}]

\textit{2}. \textit{Dynamic.} Accessibility has a dynamic nature. For example, accessibility requirements evolve, for example, a few years ago, accessibility was designed for disability, but now it is more designed for general people. Besides, during the development life cycle, a valid requirement about accessibility can turn into a set of important functionalities [\faWrench \ \textbf{S9}]. Dynamic diversity provides an important step not only towards the development of more effective interfaces for a specific group of people, but also for the more general problem of "Universal interface design" \cite{gregor2001designing}.

  \faThumbsODown \ "\textit{When presenting content, this information will be useful in explaining potential accessibility issues that could come up in widgets you choose or that might be built into the framework of the platform you are using... If you are a developer, our goal is to provide you with some idea about potential problems and issues to keep in mind and to get you started on finding ways to solve the problems. The \textbf{accessibility of dynamic content} has improved over the past several years, but there are still complications. There are many ways to create dynamic content, and thus there are many ways to make it both inaccessible and accessible.}" [\textit{I2-Programmer and Tester}]


\begin{center}
\begin{tcolorbox}[colback=gray!10,
colframe=black, 
width=5.4in,
arc = 1mm,
boxrule=0.5 pt,
]
\textbf{Finding 1}

\textbf{Accessibility is not properly integrated into \textit{general} software projects}. In our data sample, only 30$\%$ of participants have direct accessibility-related work experience. In addition, accessibility is adopted in most of their current development projects in a rather light-weight way, e.g., confined to colors and layout of interface design.

\end{tcolorbox}
\end{center}

\subsubsection{How do work characteristics impact accessibility development and design in practice}
\label{SubsecRQ1_ Work}

Almost all of the interviewees mentioned that work characteristics impact accessibility development and design in practice. In this section, we discuss what work characteristics practitioners are concerned most with regarding accessibility development and design.  We summarized three statements under this topic (see [\textbf{S10}] - [\textbf{S12}] of Table \ref{Table2_Statements}).

\textbf{Skill variety}: Interviewees identified two main differences regarding skills if they consider incorporating accessibility into their software systems. Firstly, interviewees noted that developing an application incorporating a range of accessibility supporting features presents distinct technical challenges [\textbf{S10}]. Secondly, interviewees suggested that a wider variety of skills is required for incorporating sufficient accessibility support features into a system. This can make accessibility design more challenging if a developer lacks such skills. From the interface to architecture design, accessibility is more complicated than many think -- numerous obstacles hinder more practical accessibility design and implementation [\textbf{S10}].  Around 26$\%$ of respondents agreed and strongly agreed that working practices with accessibility indicate the degree to which the projects would be successfully integrated with accessibility design. However, nine interviewees mentioned that they had not learned accessibility knowledge systematically at their universities. The paucity of related knowledge they have makes it difficult for them to carry out appropriate accessibility design and development. 

 \faThumbsODown\  "\textit{For the development a system with accessibility, developer can write the code for a particular business requirement once they gave learning a programming language, \textbf{but for accessibility design, they should learn the standards and other specific skills}. }" [\textit{I9-Programmer and Tester}]

 \faThumbsODown\  "\textit{In the context of accessibility, even small changes in the interface will hinder users to access the system. Development team and developers should be able to handle the changes and make the accessibility truly usefully for users. This is not easy, we don't know what exactly users wants. \textbf{It seems interface development takes the majority responsibilities for accessibility, which is not true and enough}. }" [\textit{I2-Programmer and Tester}]
 
  \faThumbsODown\ "\textit{It's unlikely that you'd be expected to know the ins and outs of all accessibility best practices if you aren't also playing a technical role in the project, but having a working knowledge is key}." [\textit{I12-Tester}]

\textbf{Task identification}: Interviewees reported a few differences when incorporating accessibility in terms of task identity. Two interviewees suggested one difference: it is harder to make an accurate plan for incorporating accessibility into projects [\textbf{S11}], and 34$\%$ and 17$\%$ of online survey respondents agree and strongly agree with this statement. Three interviewees noted that incorporating accessibility means they often have less control over their progress towards target completion. Once starting accessibility-related design and development, they might require (much) more effort in eliciting requirements, consulting, or end-user testing [\textbf{S10}], and 35$\%$  and 15$\%$ of online survey respondents agree and strongly agree with this statement. They mentioned that it is hard to guarantee achievement of other software quality attributes, such as performance. This is especially so for those who do not have previous work experience and accessibility relevant knowledge [\faWrench \ \textbf{S12}]. 

 \faThumbsOUp\ "\textit{Often times, companies pay for auditing, fix all issues identified, but shortly find themselves back at square one after making changes to their digital experience. They are failing to consider accessibility as an on-going initiative. If you don’t have the tools and processes in place for ensuring accessibility as you continuously improve your digital experience, accessibility issues will persist, forcing you to backtrack when push comes to shove. \textbf{This retroactive approach impedes digital innovation, increases development costs and extends delivery timelines}}." [\textit{I5-Designer}]
 
  \faThumbsODown \ "\textit{The time allotted on computers could be stressful to the body. Those with physical impairments as a matter of fact have more difficulty utilizing the web compared to normal users. People with mobility issues for example may find it difficult to move the mouse to the desired target audience. Designing therefore for low physical efforts is important to take into account}." [\textit{I7-Manager}]

\textbf{Interaction outside the organization}: Interviewees reported that incorporating accessibility into a system means they have more challenges when communicating with their clients. Development teams are not often adequately trained on appropriate techniques required to address accessibility in their roles [\textbf{S11}]. 30$\%$  of the online respondents agree and strongly agree that interacting with outside stakeholders regarding accessibility is difficult.

 \faThumbsODown\ "\textit{Integrating accessibility principles into every aspect of your system is not hard, \textbf{but takes commitment, constant communication and a sound strategy}. In this case, you need to consult experts and communicate with you clients.}" [\textit{I7-Manager}]

\begin{center}
\begin{tcolorbox}[colback=gray!10,
colframe=black, 
width=5.4in,,
arc = 1mm,
boxrule=0.5 pt,
]
\textbf{Finding 2}

Accessibility design and development in practice requires \textbf{significant specific skills and knowledge} and task management. However, most practitioners do not have such skills and knowledge. A set of concerns that developers have when incorporating accessibility into their projects, such as \textbf{task identification} and \textbf{communicatio\textbf{n} with their clients }about accessibility issues.


\end{tcolorbox}
\end{center}

\subsubsection{How do organizational factors impact accessibility design and development in practice?}
\label{SubsecRQ1_ Organizational}

Organizational factors include developmental, technological, business, operational, and social factors that impact a software system in many ways \cite{bi2018architecture} \cite{lavallee2015good}.  We discussed with interviewees the organizational factors that they found to have an impact on accessibility development and design. Company value, culture, and project leadership are other factors interviewees mentioned that strongly impact accessibility development and design in practice, and nine statements are summarized on this topic ([\textbf{S13}] - [\textbf{S21}] in Table \ref{Table2_Statements}). The results of the online survey also confirmed that organizational factors will impact accessibility development and design (i.e., six out of nine statements are scored over 4). 

Ten interviewees identified that accessibility development and design is context-dependent [\textbf{S18}]. 34$\%$ of the survey respondents identified this statement as important (i.e., agree and strongly agree). For example, the same requirement could be implemented in one project, but it is difficult to be implemented in another project because of context changes (e.g., Web App development v.s. Mobile App development). 12 interviewees noted that big companies, such as Microsoft and IBM, have more resources and experience to focus on accessibility development and design, and those companies put accessibility as a high priority [\textbf{S13}][\textbf{S14}][\textbf{S21}]. Five interviewees mentioned that some companies consider accessibility as a vital factor to extend their target market [\textbf{S18}]. In contrast, small companies focus more on functional requirements implementation and often neglect accessibility or other quality attributes. Previous studies also confirmed that large and well-known software projects have better accessibility design than small companies' products \cite{alshayban2020accessibility}. Seven interviewees mentioned that a set of models and metaphors attempt to differentiate accessibility from prominent notions of disability accommodation or technical compliance to provide a more attainable model for practitioners. However, their work has failed to sufficiently address this issue. Various practical working constraints of practitioners and projects, such as tight time-frames, lack of tools and design guidelines, lack of knowledge and experience, lack of support, and limited budgets, will impact the incorporation of accessibility success [\textbf{S15}][\faWrench \ \phone \ \faGroup \ \textbf{S16}][\textbf{S17}][\textbf{S19}].

 \faThumbsODown\ "\textit{\textbf{Supports from companies} are vital for accessibility design, unfortunately, our company does have much focus and budgets to support this functionalities}." [\textit{I3-Designer}]

 \faThumbsOUp \  "\textit{We don’t have to worry about accessibility. The framework we’re using takes care of it for us. I think this is a fairly common sentiment. After all, designing an accessible experience is difficult, especially for folks not intimately familiar with certain users’ needs. And if a piece of software says it’s “solved” the accessibility piece for a team, that’s pretty appealing}." [\textit{I10-Designer}].


\begin{center}
\begin{tcolorbox}[colback=gray!10,
colframe=black, 
width=5.4in,
arc = 1mm,
boxrule=0.5 pt,
]
\textbf{Finding 3}

\textbf{Organizational factors} impacting the ability to address accessibility in software projects include lack of appropriate group resources, available expertise, time, budget and support from companies. These strongly impact accessibility design and development in practice. 

\end{tcolorbox}
\end{center}

\subsection{RQ2: How does addressing accessibility needs fit into the software development life cycle?}

To answer RQ2, we investigated how accessibility fits into four stages, i.e., software requirement elicitation for accessibility (Section \ref{RQ2_requirement}), software design and implementation for accessibility (Section \ref{RQ2_design}), software testing and quality assurance stage for accessibility (Section \ref{RQ2_testing}), and software evaluation for accessibility (Section \ref{RQ2_evaluation}) and what challenges practitioners have.

\label{results of RQ2}

\subsubsection{Software requirement elicitation for accessibility}
\label{RQ2_requirement}

\textbf{Effort towards accessibility requirements collection}. We summarized seven statements (see [\textbf{S22}] - [\textbf{S28}] of Table \ref{Table2_Statements}) on this topic. Incomplete, inaccurate, or vague requirements are potential risks in any projects. The main challenges reported by interviewees included the time and effort needed for accessibility-related requirements collection, aligning with diverse software end-user characteristics and challenges, and involving the right set of people at the start of a project. Nearly every interviewee mentioned that incorporating accessibility requirements into a system requires significant effort [\textbf{S21}]. Interviewees also mentioned that accessibility requirements are more uncertain than detailed functional descriptions. The accessibility requirements usually include a conceptual description of the goal after incorporating accessibility-supporting features into projects. Another accessibility requirement challenge mentioned by three interviewees is that accessibility requirements are usually documented ambiguously [\textbf{S24}]. 34$\%$  of our survey respondents agree and strongly agree with this statement, and 12 survey comments confirmed that accessibility requirement elicitation is hard due to the vague requests from the clients or vague documentation. 

 \faThumbsODown\ "\textit{There’s a problem in the design world with regards to accessibility. \textbf{Accessibility requirements are often not documented clearly}, consistently, or in a way for professionals to easily follow and implement.}" [\textit{I5-Designer}]

 \faThumbsOUp\ "\textit{\textbf{By beginning to be more mindful about documenting accessibility requirements}, it can truly change the culture of design}." [\textit{I7-Manager}]

According to our interviewees, new techniques for eliciting accessibility requirements have been developed. Three interviewees mentioned that Machine Learning (ML) and AI techniques have been adopted in their companies to assist with accessibility requirement elicitation [\textbf{S27}]. For example, one interviewee mentioned they used mouse tracking techniques to elicit requirements, i.e., they will record where the users give up clicking and exit the application, and those places could be difficult for users to proceed. 23$\%$ of the survey respondents agree and strongly agree with this statement. Two interviewees mentioned that they are using a checklist for identifying accessibility requirements to analyze the environment and the problems of end-users, obtaining accessibility requirements as a priority. Three interviewees noted that identifying domain requirements is also critical, i.e., tasks that need to be supported, social and cultural dynamics, and environmental factors. Finally, it is necessary to identify key technological requirements during this phase, including the availability of hardware, software, plug-ins, and assisting technologies for the context of end-users. 

Many accessibility guidelines are available, but interviewees and survey respondents claimed that applying them is expensive. Five interviewees discussed the lack of knowledge and understanding of the principles that affect accessibility development and design in practice. They also mentioned that more efforts and time could be obtained on requirements elicitation for dealing with people with different types of disabilities and the elderly. Two interviewees mentioned that ontology was used to model the knowledge of accessibility guidelines. However, they noted that some companies are struggling to identify accessibility requirements due to tight marketing and release schedules, lack of resources, or that they are seen as not a core task [\faGroup \ \textbf{S25}][\faWrench \ \textbf{S26}][\textbf{S28}]. The survey results also confirmed those three statements, which scored 4.03, 4.00, and 3.40, respectively. 

\textbf{Strategies to address accessibility requirements}. Three interviewees said that analysis activities (user and task analysis) play an essential role. The interviewees stated that some developers simply thought of software requirements for accessibility as being only front-end design [\textbf{S23}]. For example, solutions like onscreen keyboard, screen readers, and text-to-voice basic functionalities. However, many accessibility requirements are far more complicated than that, which practitioners should be more aware of. Around 34$\%$ of the responses tend to agree and strongly agree, and this statement scored 4.05 from the online survey.

 \faThumbsODown\  "\textit{As a project manager, you're on task for connecting the dots and keeping work streams on track. Your ability and willingness to appropriately plan for accessibility, engage the right people at the right time, communicate the need and value of accessibility, and hold teams accountable for meeting accessibility \textbf{requirements will ultimately play a major role} in delivering accessible websites and apps.}" [\textit{I10-Designer and Manager}]

 \faThumbsODown\ "\textit{I would like to concentrate on 4 principles mainly ( POUR - perceivable, operable, understandable, Robust) and 12 guidelines of Text alternatives by providing alt attributes, Time based media by proving captions, Adaptable by all assistive technologies like screen readers/magnifiers, Distinguishable, Keyboard accessible.}" [\textit{I12- Designer and Programmer }]

\begin{center}
\begin{tcolorbox}[colback=gray!10,
colframe=black, 
width=5.4in,
arc = 1mm,
boxrule=0.5 pt,
]
\textbf{Finding 4}

There are some \textbf{obstacles} to collecting accessibility requirements, such as it being time-consuming, the lack of resources, and limited potential design and implementation strategies are used to fulfill accessibility requirements.

\end{tcolorbox}
\end{center}


\subsubsection{Software design and implementation for accessibility}

\label{RQ2_design}

\textbf{Architectural design}. Interviewees stated that there are differences in design and implementation if the software incorporates accessibility. Firstly,  interviewees mentioned that the high-level architecture design for incorporating accessibility into a system typically contains richer hardware design, feature engineering, and model design [\phone \ \textbf{S34}] \cite{bass2003software}. However, in many situations, a lot of companies will not consider putting additional effort into such design. As a consequence of less-design-planning, interviewees reported very little up-front thought into the software architecture to enable developers to achieve suitable accessibility solutions \cite{bi2018systematic}.

 \faThumbsODown\ "\textit{Companies that have \textbf{embraced accessibility as a component of their design}, development and quality testing processes, such as American Airlines and Chase Bank, continue to win awards for innovation and customer experience, while providing digital access to all users, regardless of abilities. Integrating accessibility principles into every aspect of your digital operations is not hard, but takes commitment, constant communication and a sound strategy}." [\textit{I10-Manager}]

Secondly, accessibility requirements need to be separately designed for low coupling and re-use [\textbf{S34}]. For instance, one interviewee mentioned that "\textit{I}\textit{ divide the accessibility functionalities into several steps for re-use and also may use existing libraries for each step.}"  Thirdly, refactoring is a downside when needing to incorporate accessibility into a system. As many systems do not consider accessibility at the beginning, but then want or need to incorporate accessibility into the system later [\faGroup \ \textbf{S33}], many problems present. Consequently, incorporating accessibility during system refactoring is more challenging and time-consuming. Similarly, such a concern was observed in a study by Horton and Sloan \cite{horton2014accessibility}. The authors point out that accessibility audits are typically performed during quality assurance phases, and user acceptance testing and solutions for the issues identified usually take place in code.

 \faThumbsODown\ "\textit{The first phase of the development lifecycle is design and UX conception. When companies consider accessibility at this stage it sets a strong foundation for their overall strategy. If website or app updates are not designed with inclusive design principles in mind, \textbf{it will undoubtedly create more obstacles later on}. The following actions will set you up for a successful approach to accessibility}." [\textit{I5-Designer}]

\textbf{Domain-dependent design}: Five interviewees mentioned that accessibility development and design based on the platforms, domains, and users [\textbf{S29}]. We found accessibility is most likely incorporated into domains such as banking, education, government, voice recognition software, transportation system, and web services domains (see Fig 2). We concluded these are the software domains  in which many of our survey participants work. A variety of system domains show a strong influence by local, national, and international legislation. More accessibility related software development could focus on other sectors, showing the importance of producing accessible software for a range of systems. 

Regarding software platforms, the majority of survey respondents focused on developing Web and Mobile applications. We also noted a trend in adopting other process-oriented frameworks to approach accessibility in software engineering. Despite the substantial uptake of mobile applications, software practitioners concerning proposals for supporting accessibility in mobile applications are still more limited in comparison to Web accessibility [\phone \ \textbf{S34}]. However, it is noteworthy that many software practitioners mentioned that techniques to support the development of accessible mobile applications had started to receive recent attention from the research community. 

\textbf{Models and standards}: Software architecture addresses the fundamental structure and tactic of the system in order to offer the desired accessibility design based on sound design decisions. Related to other kinds of support provided by the architectures, there are recurrent subjects like seeking the improvement of usability or information fusion from different resources [\textbf{S29}]. In addition, there is a clear trend in interfaces working with multi-model interfaces and systems. Regarding the communication between components/layers and other systems, we would like to highlight that most systems used typical data structures encapsulated in files, such as microservices as architecture patterns. Regarding the standards, almost every interviewee mentioned that they applied and followed a set of standards. We noted the predominance of WCAG standard, which has been established as the primary reference concerning accessibility guidelines and design issues for web applications. However, five interviewees mentioned that they have worked as web practitioners for over a decade, and they found some of the standards (e.g., WCAG) to be overwhelming and confusing. They are even more overwhelming for those who have less accessibility work experience [\textbf{S35}].


\textbf{Challenges for accessibility design and implementation}. Implementation of accessibility-supporting solutions for software is an iterative process [\faGroup \ \textbf{S33}]. Key activities include several tasks: the project manager and the work team review the project plan to achieve a common understanding and commitment to accessibility requirements. The work group sets or updates the implementation environment. In this paper, we use adoption and implementation theory and look for an empirical approach observing the actual factors that play a role in the process of accessibility implementation. Our interview and survey results show that accessibility innovations initiation and implementation models identify organizational processes of resistance and support accessibility implementation. The model contains many of the innovation-related elements identified in other models and frameworks but instead of being focused on the individuals within organizations or extending such models to include items that support or resist the initiation and implementation of software products.

There are Functional Requirement (FR) and Non-functional Requirement (NFR) categories under the first layer, and interface, control, and entity components are in the third layer under the Type-F requirements. The change and addition of requirements are tracked in this implementation. Six interviewees mentioned that accessibility often falls upon an organization's practitioners to take it upon themselves to decide whether or not to implement accessibility and how to do it. However, business pressures have been found to motivate practitioners to achieve short-term goals rather than the longer-term or indirectly profitable work of accessibility [\faWrench \ \textbf{S30}] [\textbf{S32}].

 \faThumbsODown \ "\textit{\textbf{Developers typically have the hardest time with accessibility. }They need specific expertise to know how to \textbf{resolve accessibility issues within code that is resulting in inaccessible page elements}. This requires extensive accessibility training on the techniques for following the WCAG 2.0 AA or leveraging accessibility experts for guidance on coding options.}" [\textit{I8-Tester}]

 \faThumbsODown\ "\textit{Many developers resort to using free online testing tools, but find themselves putting in\textbf{ double the work and still struggling to maintain compliance}. These tools are inconvenient and often provide false positives, which is why they ultimately are not worth it. For agile development cycles, developers need sufficient tools and streamlined technical support to efficiently maintain accessibility}." [\textit{I10-Designer and Manager}]

\begin{center}
\begin{tcolorbox}[colback=gray!10,
colframe=black, 
width=5.4in,
arc = 1mm,
boxrule=0.5 pt,
]
\textbf{Finding 5}

\textbf{Software design} for accessibility is often in current practice \textbf{short-term and mostly focusing on UI design goals and approaches}. Many projects have to be refactored over many iterations in order to incorporate accessibility solutions after the initial release.
There is a set of \textbf{barriers} for software implementation to achieve appropriate accessibility solutions, such as lack of or overwhelming design standards and guidelines, which are more difficult for less experienced developers.

\end{tcolorbox}
\end{center}


\subsubsection{Software testing and quality assurance for accessibility}
\label{RQ2_testing}

\textbf{Quality assurance deterrents}. Testing is an essential activity in the software industry as it allows one to control and improve the quality of software products. In addition, an important way to ensure the achievement of objectives in each phase of the software development is to apply a testing process \cite{sanchez2014toward}. Additional efforts and representative end-users are needed for sufficient accessibility testing to be undertaken. We summarized three statements ([\textbf{S36}] - [\textbf{S38}] in Table \ref{Table2_Statements}) from the interviews, and all statements scored by the survey respondents are over 4.00, and 61$\%$ of the respondents \textit{agree} and \textit{strongly agree} quality assurance is a vital step for accessibility development and design. As we mentioned earlier, accessibility is a quality that makes sure that end-users can access the system successfully. There are also other quality attributes that overlap with accessibility, such as security, privacy, and usability. Seven interviewees mentioned that other quality attributes are difficult to guarantee when needing to incorporate accessibility into a system [\textbf{S36}]. For example, one interviewee mentioned:

\faThumbsODown\ "\textit{We have been developing the online shopping for visual impaired people, and it is going well. They can shop using our voice assistant functionalities, and it will read everything shown in the screen. However, how to guarantee \textbf{security, performance, and privacy} is a headache}. " [\textit{I6-Programmer and Tester}]

The overall conclusion is that participants consider testing is important for accessibility, and there are a set of benefits and challenges in accessibility testing. Firstly, there is a long list of FRs and NFRs that needed to be tested. FRs testing includes hardware testing (e.g., keyboard testing) and software testing (e.g., voice reader). NFRs testing includes color usage, font, layout design, and other quality attributes [\textbf{S36}]. 

Seven interviewees also mentioned that accessibility testing is vital for different platforms, such as web and mobile app testing. There are some challenges including that developers are unlikely to write dedicated accessibility test suites, and existing test suites tend to be weak in finding many accessibility problems. A set of tools that participants mentioned can significantly reduce the manual efforts of accessibility testing [\phone \ \textbf{S38}]. These included Anywhere (a screen reader tool) and Hera (used to check the software application style). Seven interviewees mentioned that standards and checklists are helpful for accessibility testing. However, the current checklists and standards need to be updated for a more comprehensive understanding and to address new interaction technologies \cite{yusop2020revised}. Four interviewees mentioned the importance of manual accessibility testing, such as hiring disability end-users and getting feedback [\textbf{S37}].

\faThumbsODown\  "\textit{These automated programs are incredible resources. \textbf{Modern websites and web apps are complicated things that involve hundreds of states, thousands of lines of code, and complicated multi-screen interactions.} It’d be absurd to expect a human (or a team of humans) to mind all the code controlling every possible permutation of the site, to say nothing of things like regressions, software rot, and A/B tests.}" [\textit{I7-Manager}]

\faThumbsODown\ "\textit{If you’re not using a testing service that includes this rule, it won’t be reported. The code will still ``pass”, but it’s passing by omission, \textbf{not because it’s actually accessible}}." [\textit{I10-Designer and Manager}]

\faThumbsOUp\  "\textit{\textbf{Accessibility testing is no different from the other types of testing required for quality assurance} — testing a digital experience to check it is fit for a certain purpose and that the developers followed the proper guidelines and requirements.}" [\textit{I11-Manager}]

\faThumbsODown\ "\textit{In terms of tools,\textbf{ your QA team should have access to the same accessibility testing tools as your developers}. It’s important for your QA team to easily show your developers where and what issues were identified, so creating an avenue of communication between these two teams is essential for maintaining delivery timelines}." [\textit{I2-Programmer and Tester}]

\begin{center}
\begin{tcolorbox}[colback=gray!10,
colframe=black, 
width=5.4in,
arc = 1mm,
boxrule=0.5 pt,
]
\textbf{Finding 6}

A set of FRs and NFRs (e.g., privacy) are worth to be paid attention to for accessibility \textbf{testing}, and tools are helpful for automatic testing. Manual testing is essential, but it is challenging to recruit end-users and to get feedback from them.

\end{tcolorbox}
\end{center}

\subsubsection{Software evaluation for accessibility}
\label{RQ2_evaluation}

\textbf{Evaluation across stages}. To ensure accessibility as a key software feature outcome, development teams must ensure that accessibility and inclusion are embedded in their mindset, process, and tools, rather than in the head of a single accessibility expert. We summarized six statements from the interviews and their scores from the online survey (see [\textbf{S39}] - [\textbf{S44}] in Table \ref{Table2_Statements}). Five out of six statements were scored over 4. Almost all interviewees mentioned that evaluation is a \textbf{key} step for achieving required software accessibility. They gave a range of evaluation methods that assist developers in creating interactive electronic products, services, and environments that are both easy and pleasant to use by the target audience. Accessibility evaluation is a broad field that combines different disciplines and skills. It encompasses technical aspects such as the standards and guidelines [\phone \ \textbf{S44}], as well as non-technical aspects such as involvement of end-users during the evaluation process [\textbf{S43}]. Eight interviewees stated the use of iterative and user-centered evaluation is essential for accessibility design (e.g., understanding users, tasks, and contexts of projects). This involves studying existing style guidelines and standards [\phone \ \textbf{S44}]. In addition, it is better to interview end-users for understanding the products regarding strengths, weaknesses, and expectations for accessibility. However, seven interviewees also mentioned some key challenges, such as getting sufficient feedback from end-users for evaluation and the time needed to be spent to do this [\textbf{S40}] [\textbf{S41}][\textbf{S42}].

\textbf{Tool usage for accessibility evaluation}. Five interviewees mentioned that automated and semi-automated accessibility evaluation tools are essential to streamline the process of accessibility assessment, and ultimately ensure that software projects, contents, and services meet accessibility requirements. Different evaluation tools may better fit different needs and concerns. 34$\%$ of respondents (agree and strongly agree) indicated that they used a set of tools and checklists to evaluate accessibility. However, four interviewees mentioned that no single tool could determine if a project meets accessibility guidelines [\textbf{S39}]. The accessibility of a software product can be regarded as a category of properties or requirements, which ensure that it can be used in equivalent ways (considering aspects such as comfort, security, or cost) by people in the broadest range of contexts of use, especially accounting for users with disabilities. Thus, an development teams typically uses accessibility evaluation tools to deal with requirements (e.g., specification and testing).

\faThumbsODown\  "\textit{Some designers needing to meet U.S. Section 508 standards chose to provide alternative "modes of operation and information retrieval". \textbf{However, in some cases where the standard was technically met by providing an alternative, the products were awkward to use or were totally unusable by some people with disabilities.} These cases illustrate the importance of going beyond just meeting a minimum accessibility standard without sufficient evaluation.}" [\textit{I14-Programmer}]


\begin{center}
\begin{tcolorbox}[colback=gray!10,
colframe=black, 
width=5.4in,
arc = 1mm,
boxrule=0.5 pt,
]
\textbf{Finding 7}

Evaluation is a \textbf{core} step for accessibility design, and workflows and domain models are critical in evaluation activity. To sum up, evaluation varies and \textbf{heavily depends on the project domain}.

\end{tcolorbox}
\end{center}

\subsection{Gaps differ across demographic groups}
\label{Gap differ across groups}

We analyzed the statements across three pairs of demographic groups in Table \ref{Table2_Statements}, i.e., \textbf{Direct} and \textbf{Indirect} groups, \textbf{Big} and \textbf{Small} groups, and \textbf{Web} and \textbf{Mobile App} development groups. As we presented in Section \ref{survey}, we used P-value to indicate whether the differences in the agreement for each statement, and the \textit{Effect Size } is used to indicates the magnitude of difference between groups. For example, the mean score of \textbf{S1} for Direct and Indirect groups is - 0.38, which means Indirect group agrees with this statement more. Light grey color indicates the latter group is more likely to agree with the statement, and Dark grey color indicates that the former group is more likely to agree with the statement.

\textbf{Direct v.s. Indirect accessibility-related work experience groups}. There are six statements with statistically significant differences between these two groups. It is worth noting that the Direct group is more likely to agree with these four statements more, i.e., [\faWrench \ \textbf{S1}], [\faWrench \ \textbf{S8}], [\faWrench \ {\textbf{S9}}], and [\faWrench \ \textbf{S26}]. For example, they agree that accessibility needs to be addressed in all software projects and agree with the characteristics of accessibility given by interviewees. On the other hand, Indirect group agree more that their projects lack of demands on accessibility, and they also agree that accessibility is limited to front-facing design in a light-weight way (i.e., [\faWrench \ \textbf{S26}] and [\faWrench \ \textbf{S30}]).

\textbf{Big-size v.s. Small-size groups}. There are four statements with statistically significant differences between these two groups. The small size group agrees more that they lack of resources and support needed to achieve accessible software products (i.e., [\faGroup \ {\textbf{S16}}], [\faGroup \ {\textbf{S25}], and [\faGroup \ \textbf{S33}]). On the other hand, the big-size group agrees more that a more comprehensive evaluation for accessibility is needed (i.e., [\faGroup \ \textbf{S43}]).

\textbf{Web v.s. Mobile development groups}. Seven statements are with statistically significantly different answers between these two groups. It is worth noting that Mobile App practitioners are more likely to agree that the standards for accessibility in Mobile App development are unclear and time-consuming (i.e., [\phone \ \textbf{S44}] and [\phone \ \textbf{S44}]). In addition, Mobile App practitioners agree more that accessibility development and design is time-consuming and not a core task (i.e., [\phone \ \textbf{S6}] and [\phone \ \textbf{S16}]). In contrast, Web App practitioners are more focused on the accessibility basic functionalities design and testing tasks (i.e., [\phone \ \textbf{S34}] and [\phone \ \textbf{S38}]).

Overall, the results of our survey confirm some differences in interviewees' claims. Being aware on the gaps (i.e., differences) between groups would help practitioners incorporate accessibility properly into their projects in practice.

\section{Discussion}
\label{Section 5 Discussion}

We interpret the results and discuss the implications for researchers and practitioners in this section. The goal of this study was to understand how accessibility is currently incorporated into software development and how practitioners perceive accessibility issues and challenges in their projects. We discuss below the key results and implications for researchers. 

\textbf{Accessibility awareness and challenges of accessibility development}. The results of our study show that accessibility closely fits into different stages of the software life cycle, and developers need to be aware of them when they make design decisions. Despite the vast majority of online survey respondents who acknowledged the importance of accessibility, 45$\%$ of our survey respondents indicated that their projects are suffering from issues that effects accessibility development and design, such as inadequate resources and experts [\textbf{S17}][\textbf{S19}][\textbf{S27}]. The frequent occurrence of accessibility issues in software development may result in software refactoring with more efforts [\textbf{S32}]. As such, we advocate that accessibility could be treated as an FR to ensure people using software projects more efficiently. Such accessibility design drives innovation in general products, and it can be more accessible for disabled people. This could motivate researchers to spend extra efforts on accessibility management, design, and development. Our results also suggest that accessibility needs to be included in all phases of software development but achieving good accessibility solutions takes a lot of effort and time. Practitioners suggest that software development could be integrated with accessibility design elements, which can make the software to a wide range of average users, especially users with some disabilities. Future research may focus on standardizing the operationalizing the process of incorporating accessibility in software development. 

In addition, from the results of our survey, a considerable number of participants do not have any accessibility work experience. In addition, two interviewees mentioned that in their companies, accessibility is usually considered only at a late stage of development or during system refactoring and maintenance. Given the increasing demand for more accessible software \cite{vendome2019can,grundy2020human,sanchez2017method}, practitioners need to put more emphasis on accessibility understanding and the ability to appreciate trade-offs. The benefit-cost was repeatedly mentioned by interviewees, and a certain number of companies consider accessibility would be an extra cost for them to incorporate into development. However, they often find that they have to refactor/modify their systems to add accessibility functionalities afterward during the system evolution. Thus could be end up costing them more time and resources. We encourage developers to put accessibility as a first-class, upfront requirement from the very beginning of software development. We advocate that accessibility can also be considered as a an important function not only for the specifically challenged groups of people, but potentially all users who want different ways to interact and use software to suit their preferences.

\textbf{Exploring the new success criteria}. Some standards and guidelines have been widely used. However, the majority of standards have been used for decades and need updating \cite{yusop2020revised}. For example, the Web Content Accessibility Guidelines (WCAG) has been a part of digital accessibility since 1999. Technology evolves quickly, and the standards needed to catch up to ensure that people or disabilities can access today's systems. The accessibility community (e.g., Mobile Application development) has been eagerly awaiting the release of new standards. On the other hand, in our work, we advocate that accessibility can be treated as a way to ensure people can use software projects easily. At this point, we encourage researchers to craft new success and comprehensive criteria based on different accessibility levels.

\textbf{Education for accessibility knowledge}. Our work highlights the accessibility relevant skills and knowledge that are perceived important by participants \textbf{[S9][S10][S11]}. As we discussed above, industry demands for software developers with knowledge of accessibility has increased substantially in recent years. However, respondents said that there is little knowledge about the prevalence of higher education teaching accessibility relevant knowledge \cite{sanchez2014toward}. None of our interviewees said that they had learned software accessibility design and implementation approaches at universities. Software has to be designed to work with new accessibility-supporting technologies. Given this situation, we encourage more universities or higher education to increase students' knowledge of how accessibility fits within the development life cycle. In addition, as we presented in Section \ref{Section 4 Results}, accessibility would be a generalization design that drives software innovation for using products efficiently and smartly. This also requires encouraging students to focus more on accessibility design. For example, general accessibility knowledge and standards. Educators should emphasize on ability to incorporate accessibility into systems, document well [\textbf{S23}], and test and evaluate efficiently [\textbf{S36}]. 

\textbf{Organizational factors are vital for accessibility development}. Organizational factors rank the top by our interviewees that impact accessibility development and design in practice, and 54$\%$ of the survey respondents agree and strongly agree organizational factor are critical for incorporating accessibility and the quality of accessibility design.  In addition, results of our study indicates that organization drives accessibility development of their projects. Large companies tend to have more resources for addressing accessibility issues and often prioritize it. In contrast, small-size companies find it more challenging to address accessibility, reasons including lack of expertise, tight development schedule, or lack of support from  management. However, three interviewees mentioned that small steps of accessibility design can make a big difference. All organizations need to improve their awareness of the need to address accessibility during software design and development. 

\textbf{Accessibility-driven development}. One of the main goals in this study is to investigate to what extent accessibility is integrated into existing software projects. Almost every interviewee mentioned that accessibility in today's projects has a second meaning to traditional support for challenged users -- "\textit{to make the projects more easy to be used}". Some functionalities were developed to help those with disabilities, for example, voice assistant functionalities and a more natural interface. However, more and more users expect it in certain scenarios. This indicates that participants need to think more about addressing "accessibility" in their software now than in the past. 

\textbf{General recommendations for accessibility development and design in practice.} Based on the results, discussions, and implications of RQ1 and RQ2. We provide a set of general recommendations for incorporating accessibility into the projects in practice (see Table \ref{Table_Challenges}). 

1. \textbf{Organizational factors and People - RQ1}. Before making any decisions about "Accessibility": (1) Stakeholders (e.g., designers, architects, developers, testers, and clients) in a project should reach a consensus on accessibility development and design; (2) What the exact accessibility requirements are; (3) How to address these accessibility-related requirements; (4) Put accessibility requirements as a high priority to address; (5) When making decisions about accessibility, some design decisions may be invalid in the first place, and some trade-offs might exist between design decisions. This moderates efforts since not every accessibility requirement might be met; (6) Accessibility requirements need to be well-documented; and (7) Make accessibility decisions in a team. 

\definecolor{shadecolor}{rgb}{.92,  .92, .92}

\begin{table}
\small

\caption{Challenges and recommendations for incorporating accessibility in practice based on results, discussion, and implications of RQ1 and RQ2.}
\label{Table_Challenges}
\centering

\begin{tabular}{p{6cm}|p{6cm}}

\hline

 \textbf{Challenges} & \textbf{Recommendations}\\
\hline
\hline
\rowcolor{Gray}
\multicolumn{2}{l}{\textbf{Organizational}}\\
Lack of executive sponsorship. & Strong executive support.   \\

Lack of management commitment. &   Committed sponsor or leader.\\
Organizational culture factors. & Cooperative organizational culture.  \\
Organizational size too small.  &  Face-to-Face communication with customers. \\
 &   Collocation of the teams.\\
&  Facility with proper work environment.\\

\hline
\rowcolor{Gray}
\multicolumn{2}{l}{\textbf{People}}\\

 Lack of necessary skills and knowledge.  & Include team members with high competence and expertise. \\

 Lack of project management competence. &  Include team members with motivation. \\
 Lack of team work. & Knowledge in the accessibility development process. \\
 Customer relationship. &  Self-organizing teamwork. \\
\hline

\rowcolor{Gray}
\multicolumn{2}{l}{\textbf{Process}}\\
Unclear requirements. & Accessibility-oriented requirement identification.\\

 Unclear project planning. &  Accessibility oriented interactive development process.  \\
Unclear project scope. & Commutation with end-users.\\
Lack of customer role. & Regular working schedule.\\ 
Lack of customer presence. & Lack of customer presence.   \\
Unclear standard and principle. & Unclear standard and principle.\\
Unclear standard and principle. & Unclear standard and principle. \\
Inappropriateness of testing suits. &  Inappropriateness of testing suits.\\
Inappropriateness of evaluation process. & Inappropriateness of evaluation process.\\
\hline

\rowcolor{Gray}
\multicolumn{2}{l}{\textbf{Practice}}\\
 Lack of complete set of accessibility practices. & Rigorous refactoring activities. \\
Inappropriateness of technology and tools. & Well-designed documentation. \\
 &  Delivering most important feature first.  \\
 & Correct integration testing. \\
 & Appropriate technical training to team. \\
\hline

\end{tabular}
\end{table}

2. \textbf{Process and Practices - RQ2}. For accessibility development and design in the lift cycle, we listed several challenges and recommendations in Table \ref{Table_Challenges}. For example, (1) Maintaining problematic accessibility requirements is recommended; (2) Transforming accessibility requirements to other artifacts, e.g., design decisions and source code; (3) Practitioners can maintain accessibility-related elements as a team; (4) Being aware of the relationships between accessibility requirements and target users; (5) The relationship between accessibility and other types of activities can be unidirectional or bi-directional; (6) Accessibility needs to be addressed right from the early phases of software developments (e.g., requirement engineering or software design); (7) Have an iterative process to fit accessibility into the whole life cycle; (8) Use a dedicated approach (e.g., the accessibility framework) and appropriate supporting tools; (9) Consider machine learning and AI-based techniques to support accessibility design and evaluation; and (10) Generalize accessibility from dedicated design to a general design.

\section{Threats to Validity}
\label{Section 6 Threats to validation}

We use the guidelines in \cite{wohlin2012experimentation} to discuss key threats to the validity in this work.

\textbf{Construct validity} reflects to what extent the research questions and  methodology are appropriately used in a study. A threat in Stage 1 (i.e., interviews) is whether or not our interviewees are representative. To reduce this threat, we invited 15 practitioners who come from three companies in two countries. In addition, the projects they have worked on cover a wide range of domains. However, they may not be representative of all practitioners. To mitigate this potential bias, we have carefully chosen questions and topics for the interviews and performed a survey that involves a large population of practitioners from various companies around the world. Our survey respondents completed the survey based on their opinions and perception. It is possible that they conflate the skills that are very important and the skills that are very relevant to their projects or industrial contexts. To mitigate this threat, we have tried to survey a large number of practitioners. A total of 365 practitioners from various companies in 28 countries across five continents participated.

\textbf{Interval validity} focuses on factors that may influence the validity of the results. The main threat in our study is whether the data we analyzed and coded can answer our research questions. One threat exist in Stage 2 (i.e., the online survey), and it is possible that some of our survey respondents did not understand some of the statements well. To reduce this threat to validity, we provide an "I Don't Know" option in our survey, and we find that the number of respondents who choose this option to be small (i.e., 5.3$\%$). We also translated our survey to Chinese to ensure that respondents from China can understand our survey well. It is also possible that we draw the wrong conclusions about respondents’ perceptions from their comments. To minimize this threat, we read transcripts many times and checked the survey results and the corresponding comments several times.

Another thread in Stage 1 (i.e., interviews) is that the selection of statements produced at the end of the interview may not be comprehensive and may be biased to the background of experts – who may not be able to articulate their own opinions. To mitigate this bias, we have taken the following steps:
\begin{itemize}
\item Aside from asking direct questions of what opinions about accessibility they deem important, we also asked them to discuss topics that they not explicitly mentioned. The topics were selected from accessibility text books and online resources; they include concepts, comprehension, programming language, requirements, design implementation, testing, tool usage, standards, and others.
\item We have performed a survey to check whether the interviewee' opinions are perceived to be correct by a large number of practitioners.
\end{itemize}

\textbf{External validity} concerns the generality of our study results to other settings. To improve the generalizability of our results and findings, we interviewed 15 respondents from 3 companies and surveyed 365 respondents from 26 countries working for various companies (including Google, Huawei, Alibaba, Microsoft, Hengtian, and other various small to large companies) or contributing to open source projects hosted on GitHub. Still, our findings may not generalize or represent the perception of all software engineers. For example, most of the respondents are from China and the U.S. It would be interesting to perform another study to investigate more software engineers to perceive accessibility in the future. 

Another threat related to the completeness of our 44 statements about accessibility in general software projects. In this paper, we finalized these topics and statements based on the open-ended interviews of 15 interviewees. At the end of our survey, we also asked the respondents to provide additional opinions about accessibility. Among the 365 responses, 176 respondents provided comments for additional opinions. We also manually analyze these additional considerations of accessibility. Next, we applied closed card sorting to categorize the comments, i.e., we tried to categorize them into the eight topics, and we left the comments which belong to none of these topics. We notice most of the comments provided a supplementary explanation to our 44 statements. This threat could be removed by including their two statements and inviting more developers to participants.

\section{Related Work}
\label{Background_and_Related_Work}

We review a range of key related works from two perspectives: accessibility development and design in practice (see Section \ref{related work: design and development for accessibility}). We then review key related works investigating practitioners' opinions for specific topics in the software engineering field (see Section \ref{related work: developer expertise}). 

\subsection{Software design and development for accessibility}
\label{related work: design and development for accessibility}

The accessible design ensures content is faithfully rendered and can be interacted across a broad spectrum of devices, platforms, assisting technologies, and operating systems by the widest possible range of end-users \cite{persson2015universal, erlandson2007universal}. In physical environments, everyone benefits from lower curbs, automatic door openers, ramps, and other features provided for disability access. For example, a part of multimedia that allows voice narration with captioning or transcription is inaccessible to students with hearing issues. There have been some attempts to investigate accessibility in software development. The closest work to ours is Paive \textit{et al.}'s study  \cite{paiva2020accessibility}. The authors conducted a literature review on accessibility and how it fits into software engineering processes. They investigated guidelines, techniques, and methods that have been presented in the literature. The authors also provided designers and developers with updated methods that contribute to process enrichment and valuing accessibility. Their work also presented the gaps and challenges that deserve to be investigated. This work motivated us to investigate how accessibility fits into the development life cycle. However, we aim to gain \textit{participants}' feedback and opinions regarding accessibility development and design in practice. We also try to identify key gaps across demographic groups that highlight accessibility development and design challenges in specific circumstances.

Alshayban \textit{et al}. conducted an empirical study aiming at understanding the accessibility of Android apps \cite{alshayban2020accessibility}. The authors reported a large-scale analysis on the prevalence of a wide variety of accessibility issues in over 1,000 Android apps across 33 different application categories, and 11 types of accessibility issues were identified, for example, TextContrast and SpeakableText. The authors also presented the findings of a survey involving 66 practitioners that reveal the current practices and challenges in Android apps regarding accessibility. The results of the survey show that developers are generally unaware of the accessibility principles and the existing analysis tools are not sufficiently sophisticated to be used. In our work, we interviewed 15 practitioners and got 365 survey responses about accessibility development and design in practice. However, we focus on Mobile app development and many other general software development scenarios in terms of accessibility. Bai \textit{et al}. presented an evaluation of nine accessibility testing methods applied in the agile software process, and during the evaluation, the authors investigated different accessibility testing methods and discussed the benefit and cost of the each recommended method \cite{bai2017cost}. Nganji \textit{et al.} proposed a disability-aware software engineering process model that considers the needs of people with disabilities, hence improving accessibility and usability of the designed system. Shinohara \textit{et al.} \cite{shinohara2018teaches} conducted a survey to investigate the prevalence of higher education teaching about accessibility or faculty's perceived barriers to teaching accessibility. The results show that teaching accessibility is prevalent but shallow among U.S faculty as broad. The authors also reported that the most critical barriers to clear and discipline-specific accessibility learning objectives are the lack of faculty knowledge about accessibility.  Mealin and Murphy-Hill conducted an exploratory empirical study, which interviewed eight blind software developers to identify what challenges they are facing in software development \cite{mealin2012exploratory}. The results show that a set of accessibility challenges that developers have met, such as inaccessibility of IDEs. Their work shows that blind software developers need accessibility relevant software development tool supports.

\subsection{Studies on developer expertise}
\label{related work: developer expertise}
Prior studies explored developers' opinions about various software activities. For example, Xia \textit{et al.} conducted an empirical study to understand coding proficiency, and the authors interviewed and surveyed software practitioners. The authors identified 38 coding proficiency skills, which can be grouped into nine categories. The authors also highlighted 21 critical skills, along with the rationale given by participants. The finding could help practitioners being aware of acquired coding skills \cite{xia2019practitioners}. Murphy-Hill \textit{et al.} presented a study with 14 interviewees and 364 survey responses for understanding substantial differences between video game development and other software development. The authors contributed a new empirical foundation on which to understand those differences \cite{murphy2014cowboys}. Wan \textit{et al.} performed a mixture of quality and quantitative studies with 14 interviews and 342 survey responses from 26 countries to elicit significant differences between the development of machine learning systems and the development of non-machine learning systems \cite{wan2019does}. In our previous work \cite{9263357}, we investigated participants' opinions on release note production and usage in practice and what gaps exist between release note producers and users. 

Similarly, in this work, we followed the same approach (i.e., a mixed-methods approach approach) to investigate accessibility development and design challenges, issues, and benefits from practitioners' opinions, which help development teams incorporate accessibility in practice by highlighting essential skills to acquire.

\section{Conclusion}
\label{Section 7 Conclusion}

We investigated how accessibility fits into the development life cycle of existing software projects, and how practitioners perceive accessibility development and design in their projects.  We presented empirical evidence from participants' viewpoints highlighting the key obstacles in accessibility design and implementation that they currently face (see Table \ref{Table2_Statements} and Table \ref{Table_Challenges}).  Through our analysis, we grouped practitioner statements about accessibility issues into SE topics throughout the software development life cycle. This can help both researchers and practitioners better understand what challenges and benefits to incorporate accessibility into software projects. We advocate that accessibility should be treated as a first class consideration throughout software development. Base on the results of interviews and the online survey, we outline a set of challenges of accessibility and recommendations on how to support practitioners to consider accessibility in software development and design (see Table \ref{Table_Challenges}). With the findings, we conjecture that accessibility still needs considerable efforts to be further explored from various perspectives (e.g., the end-users). Our study also highlights opportunities for researchers and participants to address the existing issues.

\section*{Acknowledgment}
The authors would like to thank all the interview participants and survey respondents; without them, this work will never be accomplished.




\bibliographystyle{ACM-Reference-Format}
\bibliography{ref}

%

\end{document}